\documentstyle[12pt]{article}

\newcommand{\newc}{\newcommand} 
\newc{\ra}{\rightarrow} 
\newc{\lra}{\leftrightarrow} 
\newc{\beq}{\begin{equation}} 
\newc{\eeq}{\end{equation}} 
\newc{\barr}{\begin{eqnarray}} 
\newc{\earr}{\end{eqnarray}} 

\begin{document} 
\begin{center}
{\bf The Directional Rate and the Modulation Effect for
Direct Supersymmetric Matter Detection. }

\vspace{5mm} 

J. D. VERGADOS 

\vspace{3mm} 

Theoretical Physics Division, University of Ioannina, GR-45110, Greece. 
\end{center}

\vspace{3mm} 
\abstract{\it The detection of the theoretically expected dark matter
is central to particle physics and cosmology. Current fashionable supersymmetric
models provide a natural dark matter candidate which is the lightest
supersymmetric particle (LSP). Such models combined with fairly well 
understood physics like
the quark substructure of the nucleon and the nuclear structure (form factor 
and/or spin response function), permit the evaluation of
the event rate for LSP-nucleus elastic scattering. The thus obtained event rates
are, however, very low or even undetectable.
 So it is imperative to exploit the modulation effect, i.e. the dependence of
the event rate on  the earth's annual motion.  Also it is useful to consider
the directional rate, i.e its dependence on the direction of the recoiling
nucleus. In this paper we study
such a modulation effect both in non directional and directional experiments.
We calculate both the differential and the total rates using both isothermal, 
symmetric as well
as only axially asymmetric, and non isothermal, due to caustic rings, velocity
distributions. We consider We find that in the symmetric case 
the modulation amplitude is small. The same is true for the case of caustic
 rings.  The inclusion of asymmetry, with a realistic enhanced velocity 
dispersion  in the galactocentric direction, yields an enhanced
modulation effect, especially in directional experiments.}
\section{Introduction}
In recent years the consideration of exotic dark matter has become necessary
in order to close the Universe \cite {KTP}$^,$ \cite{Jungm}. Furthermore in
in order to understand the large scale structure of the universe 
it has become necessary to consider matter
made up of particles which were 
non-relativistic at the time of freeze out. This is  the cold dark 
matter component (CDM). The COBE data ~\cite{COBE} suggest that CDM
is at least $60\%$ ~\cite {GAW}. On the other hand during the last few years
evidence has appeared from two different teams,
the High-z Supernova Search Team \cite {HSST} and the
Supernova Cosmology Project  ~\cite {SPF} $^,$~\cite {SCP} 
 which suggests that the Universe may be dominated by 
the  cosmological constant $\Lambda$r.
As a matter of fact recent data the situation can be adequately
described by  a barionic component $\Omega_B=0.1$ along with the exotic 
components $\Omega _{CDM}= 0.3$ and $\Omega _{\Lambda}= 0.6$.
In another analysis Turner \cite {Turner} gives 
$\Omega_{m}= \Omega _{CDM}+ \Omega _B=0.4$.
Since the non exotic component cannot exceed $40\%$ of the CDM 
~\cite{Jungm}$^,$~\cite {Benne}, there is room for the exotic WIMP's 
(Weakly  Interacting Massive Particles).
  In fact the DAMA experiment ~\cite {BERNA2} 
has claimed the observation of one signal in direct detection of a WIMP, which
with better statistics has subsequently been interpreted as a modulation signal
~\cite{BERNA1}.

The above developments are in line with particle physics considerations. Thus,
in the currently favored supersymmetric (SUSY)
extensions of the standard model, the most natural WIMP candidate is the LSP,
i.e. the lightest supersymmetric particle. In the most favored scenarios the
LSP can be simply described as a Majorana fermion, a linear 
combination of the neutral components of the gauginos and Higgsinos
\cite{Jungm}$^,$~\cite{JDV}$^-$\cite{Hab-Ka}. 

 Since this particle is expected to be very massive, $m_{\chi} \geq 30 GeV$, and
extremely non relativistic with average kinetic energy $T \leq 100 KeV$,
it can be directly detected ~\cite{JDV}$^-$\cite{KVprd} mainly via the recoiling
of a nucleus (A,Z) in the elastic scattering process:
\begin{equation}
\chi \, +\, (A,Z) \, \to \, \chi \,  + \, (A,Z)^* 
\end{equation}
($\chi$ denotes the LSP). In order to compute the event rate one needs
the following ingredients:

1) An effective Lagrangian at the elementary particle 
(quark) level obtained in the framework of supersymmetry as described 
in Refs.~\cite{Jungm}, Bottino {\it et al.} \cite{ref2} and \cite{Hab-Ka}.

2) A procedure in going from the quark to the nucleon level, i.e. a quark 
model for the nucleon. The results depend crucially on the content of the
nucleon in quarks other than u and d. This is particularly true for the scalar
couplings as well as the isoscalar axial coupling ~\cite{Dree,Chen}.

3) Compute the relevant nuclear matrix elements~\cite{Ress}$^-$\cite{DIVA00}
using as reliable as possible many body nuclear wave functions.
By putting as accurate nuclear physics input as possible, 
one will be able to constrain the SUSY parameters as much as possible.
The situation is a bit simpler in the case of the scalar coupling, in which
case one only needs the nuclear form factor.

Since the obtained rates are very low, one would like to be able to exploit the
modulation of the event rates due to the earth's
revolution around the sun. To this end one adopts a folding procedure
assuming some distribution~\cite{Jungm,SIKIVIE,Druk} of velocities for the LSP.
One also would like to know the directional rates, by observing the 
nucleus in a certain direction, which correlate with the motion of the
sun around the center of the galaxy.

 The purpose of our present review is to focus on this last point 
along the lines suggested by our recent work \cite{JDV99,JDV99b}.
For the reader's convenience, however, we will give a description  in sects 2,
and 4 of the basic SUSY ingredients needed to calculate LSP-nucleus scattering
 cross section. We will
not, however, elaborate on how one gets the needed parameters from
supersymmetry. The calculation of these parameters  has become pretty standard.
 One starts with   
representative input in the restricted SUSY parameter space as described in
the literature, e.g. Bottino {\it et al.} ~\cite{ref2}, 
Kane {\it et al.}, Castano {\it et al.} and Arnowitt {\it et al.} \cite {ref3}.
 Our own SUSY input parameters will appear elsewhere \cite {Gomez}

After this we will specialize our study in the case of the nucleus $^{127}I$, 
which is one of the most popular targets \cite{BERNA2}$^,$\cite{Smith}
\cite{Primack}$^-$ \cite {KVdubna}.
To this end we will consider velocity distributions both isothermal,
  symmetric Maxwell-Boltzmann 
\cite {Jungm} as well as only axially asymmetric, like that 
of Drukier \cite {Druk} and non isothermal as well.
In the non isothermal case we will consider the relatively simple case
in which dark matter accumulated  as late in-fall of such  
matter into our galaxy, i.e. the Sikivie caustic rings \cite {SIKIVIE}.

 Since the expected rates are extremely low or even undetectable
with present techniques, one would like to exploit the characteristic
signatures provided by the reaction. Such are: a) The modulation
 effect, i.e the dependence of the event rate on the velocity of
the Earth and b) The directional event rate, which depends on the
 velocity of the sun around the galaxy as well as the the velocity
of the Earth. The latter effect, recognized sometime ago 
\cite {Sperg96}
has recently begun to appear feasible by the planned UKDMC experiment
\cite {UKDMC}.  We will study both of these effects in the present
work.

 In all calculations we will, of course, include an appropriate 
nuclear form factor and take into account
the influence on the rates of the detector energy cut off.
We will present our results a function of the
LSP mass, $m_{\chi}$, in a way
which can be easily understood by the experimentalists.


\bigskip
\section{The Nature of the LSP}
\bigskip

Before proceeding with the
construction of the effective Lagrangian we will briefly discuss 
the nature of the
lightest supersymmetric particle (LSP) focusing on those ingredients
which are of interest to dark matter.

In currently favorable supergravity models the LSP is a linear
combination~\cite{Jungm,JDV} of the neutral four fermions 
${\tilde B}, {\tilde W}_3, {\tilde H}_1$ and ${\tilde H}_2$ 
which are the supersymmetric partners of the gauge bosons $B_\mu$ and
$W^3_\mu$ and the Higgs scalars
$H_1$ and $H_2$. Admixtures of s-neutrinos are expected to be negligible.

In the above basis the mass-matrix takes the form~\cite{Jungm,Hab-Ka} 
\beq
  \left(\begin{array}{cccc}M_1 & 0 & -m_z c_\beta s_w & m_z s_\beta s_w \\
 0 & M_2 & m_z c_\beta c_w & -m_z s_\beta c_w \\
-m_z c_\beta s_w & m_z c_\beta c_w & 0 & -\mu \\
m_z s_\beta s_w & -m_z c_\beta c_w & -\mu & 0 \end{array}\right)
\label{eq:eg 4}
\eeq

In the above expressions $c_W = cos \theta_W$,  
$s_W = sin\theta_W$, $c_\beta = cos\beta$, $s_\beta = sin\beta$,
where $tan\beta =\langle \upsilon_2\rangle/\langle\upsilon_1\rangle$ is the
ratio of the vacuum expectation values of the Higgs scalars $H_2$ and
$H_1$. $\mu$ is a dimensionful coupling constant which is not specified by the
theory (not even its sign). The parameters $tan \beta , M_1, M_2, \mu$ are
determined by the procedure of Kane {it et al} and Castano {\it et al} in
Ref.~\cite{ref3} using universal
masses of the GUT scale.

By diagonalizing the above matrix we obtain a set of eigenvalues $m_j$ and
the diagonalizing matrix $C_{ij}$ as follows

\beq
\left(\begin{array}{c}{\tilde B}_R\\
{\tilde W}_{3R}\\
{\tilde H}_{1R} \\
 {\tilde H}_{2R}\end{array}\right) = (C^R_{ij})
\left(\begin{array}{c} \chi_{1R} \\
\chi_{2R} \\   \chi_{3R}\\
\chi_{4R}\end{array}\right)  \qquad
\left(\begin{array}{c} {\tilde B}_L \\
{\tilde W}_{2L} \\   {\tilde H}_{1L} \\
{\tilde H}_{2L} \end{array}\right) =  \left(C_{ij} \right)
\left(\begin{array}{c} \chi_{1L} \\
\chi_{2L} \\   \chi_{3L} \\
\chi_{4L} \end{array}\right) 
\label{eq:eg.8}
\eeq

\noindent
with $C^R_{ij}=C^*_{ij}e^{i \lambda_j}$
The phases are $\lambda_i=0, \pi$ depending on the sign of the eigemass.

Another possibility to express the above results in photino-zino 
basis ${\tilde \gamma}, {\tilde Z}$ via
\barr
{\tilde W}_3 &=& sin \theta_W {\tilde \gamma}
 -cos \theta_W {\tilde Z} \nonumber \\
{\tilde B}_0 &=& cos \theta_W {\tilde \gamma}
 +sin \theta_W {\tilde Z} \label{eq:eg9}
\earr
In the absence of supersymmetry breaking $(M_1=M_2=M$ and $\mu=0)$ the
photino is one of the eigenstates with mass $M$. One of the remaining
eigenstates has a zero eigenvalue and is a linear combination of ${\tilde
H}_1$ and ${\tilde H}_2$ with mixing  $sin \beta$. In the presence
of SUSY breaking terms the ${\tilde B}, {\tilde W}_3$ basis is superior
since the lowest eigenstate $\chi_1$ or LSP is primarily ${\tilde B}$. From
our point of view the most important parameters are the mass $m_x$ of LSP
and the mixings $C_{j1}, j=1,2,3,4$ which yield the $\chi_1$ content of the
initial basis states. These parameters which are relevant here
are shown in Table 1.
\begin{table}[t]  
\caption{  
The essential parameters describing the LSP and Higgs.
For the definitions see the text.
}

\vspace{0.4cm}
\begin{center}
\begin{tabular}{|l|lllllllll|}
\hline
\hline
 & & & & & & & & &  \\
 Solution & $\#$1 & $\#$2 & $\#$3 & $\#$4 & $\#$5 & $\#$6 & $\#$7 & $\#$8 
 & $\#$9 \\
\hline
 & & & & & & & & &  \\
$m_x \,(GeV)$ & 126 & 27 & 102 & 80 & 124 & 58 & 34 & 35 & 50 \\
$m_h$ & 116.0& 110.2& 113.2& 124.0& 121.0& 105.0& 103.0 & 92.0& 111.0 \\
$m_H$ & 345.6& 327.0& 326.6& 595.0& 567.0& 501.0& 184.0& 228.0& 234.0 \\
$m_A$ & 345.0& 305.0& 324.0& 594.0& 563.0& 497.0& 179.0& 207.0& 230.0 \\
tan2$\alpha$& 0.245& 6.265& 0.525& 0.410& 0.929& 0.935& 0.843& 1.549& 0.612 \\
tan$\beta$& 10.0& 1.5&  5.0&   5.4&   2.7&   2.7&   5.2&   2.6&   5.3 \\
\hline
\hline
\end{tabular}
\end{center}
\end{table}
We are now in a position to find the interaction of $\chi_1$ with matter. 
We distinguish three possibilities involving Z-exchange, s-quark exchange and
Higgs exchange.

\bigskip
\section{The Feynman Diagrams Entering the Direct Detection of LSP.} 
\bigskip

 The diagrams involve Z-exchange, s-quark exchange and Higgs exchange.

\subsection{The Z-exchange contribution.}
\bigskip

This can arise from the interaction of Higgsinos with $Z$ which can be read
from Eq. C86 of Ref.~\cite{Hab-Ka} 
\beq
{\it L} = \frac{g}{cos \theta_W} \frac{1}{4} [{\tilde H}_{1R}
\gamma_{\mu}{\tilde H}_{1R} -{\tilde H}_{1L}\gamma_{\mu} {\tilde H}_{1L} -
({\tilde H}_{2R}\gamma_{\mu}{\tilde H}_{2R}  -{\tilde H}_{2L}\gamma_{\mu}{\tilde
H}_{2L})]Z^{\mu}
 \label{eq:eg 10}
\eeq
Using Eq. (\ref{eq:eg.8}) and the fact that for Majorana particles 
${\bar \chi} \gamma_{\mu} \chi = 0$, we obtain 
\beq
{\it L} = \frac {g}{cos \theta_W} \frac {1}{4} (|C_{31}|^2
-|C_{41}|^2) {\bar \chi}_1\gamma_{\mu} \gamma_5 \chi_1 Z^{\mu}
 \label{eq:eg 11}
\eeq
which leads to the effective 4-fermion interaction 
\beq
{\it L}_{eff} = \frac {g}{cos \theta_W} \frac {1}{4} 2(|C_{31}|^2
-|C_{41}|^2)  (- \frac {g}{2cos \theta_W} \frac {1}{q^2 -m^2_Z}
 {\bar \chi}_1\gamma^{\mu} \gamma_5 \chi_1)J^Z_\mu
 \label{eq:eg 12}
\eeq
where the extra factor of 2 comes from the Majorana nature of
$\chi_1$. The neutral hadronic current $J^Z_\lambda$ is given by
\beq
J^Z_{\lambda} = - {\bar q} \gamma_{\lambda} \{ \frac {1}{3} sin^2 \theta_W -
\Big[ \,\frac {1}{2} (1-\gamma_5) - sin^2 \theta_W \Big]\tau_3 \} q  
  \label{eq:eg 13}
\eeq
at the nucleon level it can be written as
\beq
{\tilde J}_{\lambda}^Z = -{\bar N} \gamma_{\lambda} \{ \, sin^2 \theta_W 
-g_V (\frac{1}{2} - sin^2\theta_W)\tau_3 + \frac{1}{2} g_A \gamma_5 \tau_3
\} N
  \label{eq:eg 14}
\eeq
Thus we can write
\beq
{\it L}_{eff} = - \frac {G_F}{\sqrt 2} ({\bar \chi}_1 \gamma^{\lambda}
\gamma^5 \chi_1) J_{\lambda}(Z)
 \label{eq:eg 15}
\eeq 
where
\beq
J_{\lambda}(Z) = {\bar N} \gamma_{\lambda} [f^0_V(Z) + f^1_V(Z) \tau_3
+  f^0_A(Z) \gamma_5 + f^1_A(Z) \gamma_5  \tau_3] N
\label{eq:eg 16}
\eeq
and
\barr
f^0_V(Z) &=& 2(|C_{31}|^2 -|C_{41}|^2) \frac {m^2_Z}{m^2_Z - q^2} sin^2
\theta_W \\ 
f^1_V(Z) &=& - 2(|C_{31}|^2 -|C_{41}|^2) \frac
{m^2_Z}{m^2_Z - q^2}g_V (\frac {1}{2} - sin^2 \theta_W) \\
f^0_A (Z) &=& 0  \\
f^1_A (Z) &=&  2(|C_{31}|^2 -|C_{41}|^2) \frac {m^2_Z}{m^2_Z - q^2} \,\, \frac
{1}{2} g_A \label{eq:eg 13d}
\earr
with $g_V=1.0,  g_A = 1.24$. We can easily see that
\beq
f^1_{V}(Z)/ f^0_{V}(Z) = -g_V ( \frac {1}{2sin^2 \theta_W} - 1 ) \simeq
- 1.15  \nonumber
\eeq
Note that the suppression of this Z-exchange interaction compared to 
the ordinary
neutral current interactions arises from the smallness of the mixings
$C_{31}$ and $C_{41}$, a consequence of the fact that the Higgsinos are
normally quite a bit heavier than the gauginos.  Furthermore, the two
Higgsinos tend to cancel each other.

\bigskip
\subsection{The $s$-quark Mediated Interaction }
\bigskip

The other interesting possibility arises from the other two components of
$\chi_1$, namely ${\tilde B}$ and ${\tilde W}_3$. Their corresponding
couplings to $s$-quarks can be read from the appendix C4 of Ref.~\cite{Hab-Ka}
They are
\barr
{\it L}_{eff} &=& -g \sqrt {2} \{{\bar q}_L [T_3 {\tilde W}_{3R} 
- tan \theta_W (T_3 -Q) {\tilde B}_R ] {\tilde q}_L \nonumber \\
&-& tan  \theta_W {\bar q}_R Q {\tilde B}_L {\tilde q}_R\} + HC
 \label{eq:eg 17}
\earr
where ${\tilde q}$ are the scalar quarks (SUSY partners of quarks). A
summation over all quark flavors is understood. Using Eq. (\ref{eq:eg.8}) we
can write the above equation in the $\chi_i$ basis. Of interest to us here
is the part
\barr 
{\it L}_{eff} &=& g \sqrt {2} \{(tan \theta_W (T_3 -Q) C^R_{11} -
T_3 C^R_{21}) {\tilde q}_L \chi_{1R} {\tilde q}_L \nonumber \\
&+& tan  \theta_W C_{11} Q {\bar q}_R \chi_{1L} {\tilde q}_R\} 
\label{eq:eg.18}
\earr
The above interaction is almost diagonal in the quark flavor. There exists,
however, mixing between the s-quarks ${\tilde q}_L$ and ${\tilde q}_R$
(of the same flavor) i.e.
\begin{equation}
{\tilde q}_L = cos \theta_{{\tilde q}}{\tilde q}_1 
+ sin \theta_{{\tilde q}}{\tilde q}_2   
\end{equation}
\begin{equation}
{\tilde q}_R = -sin \theta_{{\tilde q}}{\tilde q}_1 
+ cos \theta_{{\tilde q}}{\tilde q}_2   
\end{equation}
with
\begin{equation}
tan 2\theta_{{\tilde u}} =  \frac {m_u(A+\mu cot \beta)} 
{m^2_{u_L} -m^2_{{\tilde u}_R} + m^2_z cos2 \beta/2}
\end{equation}
\begin{equation}
tan 2\theta_{{\tilde d}} =  \frac {m_d(A+\mu tan \beta)} 
{m^2_{d_R} -m^2_{{\tilde d}_R} + m^2_Z cos2 \beta/2}
\end{equation}
Thus Eq. (\ref{eq:eg.18}) becomes
\barr
{\it L}_{eff} &=& g \sqrt{2} \left\{ [ B_L cos\theta_{{\tilde q}} \right.
{\bar q}_L \chi_{1R} -B_R sin\theta_{{\tilde q}}{\bar q}_R \chi_{1L}]
{\tilde q}_1 \nonumber \\
&+& [ B_L sin \theta_{{\tilde q}} {\bar q}_L \chi_{1R} + 
B_R cos\theta_{{\tilde q}} {\bar q}_R \chi_{1L}] \left.
{\tilde q}_2 \right \} \nonumber   
\earr
with
\begin{center}
$B_L(q) = -\frac{1}{6} C^R_{11}tan\theta_{\omega}-\frac{1}{2} C^R_{21},
\,\,\,  q=u \ \ (charge \ \ 2/3)$   
\end{center}
\begin{center}
$B_L(q) = -\frac{1}{6} C^R_{11} tan\theta_{\omega} + \frac{1}{2} C^R_{21},
\ \ \ q=d \ \ (charge \ \ -1/3)$   
\end{center}
\begin{center}
$B_R(q) = \frac {2}{3} tan \theta_{\omega}  C_{11}, \ \ \  q=u \ \ (charge \ \ 2/3)$   
\end{center}
\begin{center}
$B_R(q) = - \frac {1}{3} tan \theta_{\omega}  C_{11}, \ \ \  q=d\ \ (charge \ \ -1/3)$   
\end{center}
The effective four fermion interaction takes the form
\barr
\lefteqn{{\it L}_{eff} = (g \sqrt{2})^2  \{(B_L cos \theta_{\tilde q} {\bar q}_L
\chi_{1R} - B_R sin\theta_{\tilde q} {\bar q}_R \chi_{1L}) }      
 & & \nonumber \\
& & \frac{1}{q^2-m_{ {\tilde q}_1^2}} (B_L cos \theta_q {\bar \chi}_{1R} q_L
 - B_R sin\theta_{{\tilde q}} {\bar \chi}_{1L} q_R) \nonumber \\
 & & + (B_L sin \theta_q q_L
\chi_{1R} + cos\theta_{\tilde q} {\bar q}_R \chi_{1L}) \frac{1}{q^2-
m _{{\tilde q}_2^2}} \nonumber \\
 & &(B_L sin \theta_q {\bar \chi}_{1R} q_L
 + B_R cos\theta_{\tilde q} {\bar \chi}_{1L} q_R) \} \label{eq:eg 18}
\earr

The above effective interaction can be written as
\begin{equation}
{\it L}_{eff} = {\it L}_{eff}^{LL+RR} + {\it L}_{eff}^{LR}
\end{equation}
The first term involves quarks of the same chirality and is not much effected
by the mixing (provided that it is small). The second term involves quarks 
of opposite chirality and is proportional to the s-quark mixing.

\bigskip
i) The part ${\it L}_{eff}^{LL+RR}$\\

Employing a Fierz transformation ${\it L}_{eff}^{LL+RR}$ can be cast in the more
convenient form
 \barr 
{\it L}_{eff}^{LL+RR} =&& (g \sqrt{2})^2  2(-\frac{1}{2})\{ |B_L|^2 
\nonumber  \\
&& (\frac{cos^2 \theta_{\tilde q}}{q^2-m_{ {\tilde q}_1^2}} + 
\frac{sin^2\theta_{\tilde q}}{q^2-m_{ {\tilde q}_2^2}})
 {\bar q}_L \gamma_\lambda q_L
\chi_{1R} \gamma^\lambda \chi_{1R} 
\nonumber  \\
&& + |B_R|^2 (\frac {sin^2 \theta_{\tilde q}}{q^2-m_{ {\tilde q}_1^2}} + 
\frac {cos^2 {\theta_{\tilde q}}} {q^2-m_{{\tilde q}_2^2}} )
 {\bar q}_R \gamma_\lambda q_R
\chi_{1L} \gamma^\lambda \chi_{1L} \} 
 \label{eq:eg.19}
\earr
The factor of 2 comes from the Majorana nature of LSP and the (-1/2) comes
from the Fierz transformation. Equation (\ref{eq:eg.19}) can be
written more compactly as
\barr
{\it L}_{eff} & = &  - \frac {G_F} {\sqrt {2}} 2\{ {\bar q} \gamma_\lambda
(\beta_{0R} +\beta_{3R} \tau_3) (1+ \gamma_5) q
\nonumber \\
 & - & {\bar q} \gamma_\lambda (\beta_{0L} +\beta_{3L} \tau_3)(1-\gamma_5)
 q \}({\bar \chi}_1 \gamma^\lambda \gamma^5 \chi_1 \}
 \label{eq:eg 21}
\earr
with
\barr
\beta_{0R} &=& \Big( \frac {4} {9} \chi^2_{{\tilde u}_R} 
+\frac {1} {9} \chi^2_{{\tilde d}_R}\Big) |C_{11} tan \theta_W|^2\nonumber \\
\beta_{3R} &=& \Big( \frac {4} {9} \chi^2_{{\tilde u}_R} 
-\frac {1} {9} \chi^2_{{\tilde d}_R} \Big) |C_{11} tan \theta_W|^2
\label{eq:eg 22}\\ 
\beta_{0L} &=& | \frac {1} {6} C^R_{11} tan\theta_W
+\frac{1}{2} C^R_{21}|^2 \chi^2_{{\tilde u}_L} + | \frac {1} {6} C^R_{11}
tan\theta_W  - \frac{1}{2} C^R_{21}|^2 \chi^2_{{\tilde d}_L} \nonumber \\
\beta_{3L} &=& | \frac {1} {6} C^R_{11} tan\theta_W +\frac{1}{2} C^R_{21}|^2
\chi^2_{{\tilde u}_L} - | \frac {1} {6} C^R_{11} tan\theta_W 
-\frac{1}{2} C^R_{21}|^2 \chi^2_{{\tilde d}_L} \nonumber 
\earr
with
\barr
\chi^2_{qL} &=& c^2_{\tilde q} \frac {m_W^2}{m_{{\tilde q}^2_1}-q^2} +
 s^2_{{\tilde q}} \frac {m_W^2}{m_{{\tilde q}^2_2}-q^2} \nonumber \\
\chi^2_{qR} &=& s^2_{\tilde q} \frac{m_W^2}{m_{{\tilde q}^2_1}-q^2} +
 c^2_{\tilde q} \frac{m_W^2}{m_{{\tilde q}^2_2}-q^2} \nonumber \\
c_{\tilde q} &=& cos\theta_{\tilde q}, \,\ s_{\tilde q}=sin\theta_{\tilde q} 
 \label{eq:eg 23}
\earr
The above parameters are functions of the four-momentum transfer which
in our case is negligible. Proceeding as in Sec. 2.2.1 we can obtain the
effective Lagrangian at the nucleon level as
\beq
{\it L}_{eff}^{LL+RR} = - \frac {G_F}{\sqrt 2} ({\bar \chi}_1 \gamma^{\lambda}
\gamma^5 \chi_1) J_{\lambda} ({\tilde q})
 \label{eq:eg.24}
\eeq
\beq
J_{\lambda}({\tilde q}) = {\bar N} \gamma_{\lambda} \{f^0_V({\tilde q}) + f^1_V
({\tilde q}) \tau_3 + f^0_A({\tilde q}) \gamma_5 + f^1_A({\tilde q}) \gamma_5
\tau_3) N
  \label{eq:eg.25}
\eeq
with
\barr
 f^0_V = 6(\beta_{0R}-\beta_{0L}) , \qquad f^1_V = 2
(\beta_{3R}-\beta_{3L})
\nonumber \\
f^0_A = 2g^0_a g_V (\beta_{0R}+\beta_{0L}), \qquad
f^1_A = 2g_A(\beta_{3R}+\beta_{3L})
\label{eq:eg 25a}
\earr
 with $g_v=1.0$ and $g_A=1.25$. The quantity $g^0_A$ depends on the
 quark model for the nucleon. It can be anywhere between 0.12 and 1.00.

We should note that this interaction is more suppressed than the ordinary
weak interaction by the fact that the masses of the s-quarks are usually
larger than that of the gauge boson $Z^0$. In the limit in which the LSP 
is a pure bino ($C_{11} = 1,  C_{21} = 0$) we obtain
\barr
\beta_{0R} &=& tan^2 \theta_W \Big( \frac{4}{9} \chi^2_{u_R} 
+\frac{1} {9} \chi^2_{{\tilde d}_R}\Big) \nonumber \\
\beta_{3R} &=& tan^2 \theta_W \Big( \frac{4} {9} \chi^2_{u_R} 
-\frac {1} {9} \chi^2_{{\tilde d}_R} \Big) 
\nonumber  \\
\beta_{0L} &=&  \frac{tan^2 \theta_W} {36} (\chi^2_{{\tilde u}_L} + 
 \chi^2_{{\tilde d}_L}) \nonumber \\
\beta_{3L} &=&  \frac{tan^2 \theta_W} {36}
(\chi^2_{{\tilde u}_L} -  \chi^2_{{\tilde d}_L})
\label{eq:eg 24}
\earr
Assuming further that $\chi_{{\tilde u}_R} = \chi_{{\tilde d}_R} 
= \chi_{{\tilde u}_L} = \chi_{{\tilde d}_L}$ we obtain
\barr
 f^1_V ({\tilde q}) / f^0_V({\tilde q}) &\simeq& + \frac{2}{9}
\nonumber \\
  f^1_A ({\tilde q})/ f^0_A({\tilde q}) &\simeq& + \frac{6}{11}
 \label{eq:eg 25}
\earr

If, on the other hand, the LSP were the photino ($C_{11} = cos\theta_W,
C_{21} = sin \theta_W, C_{31} = C_{41} = 0$) and the s-quarks were
degenerate there would be no coherent contribution ($f^0_V = 0$ if
$\beta_{0L} =\beta_{0R}$).

\bigskip
ii) ${\it L}_{eff}^{LR}$\\

From Eq. (\ref{eq:eg 18}) we obtain

$$
{\it L}_{eff}^{LR} = - (g \sqrt{2})^2 sin2\theta_{\tilde q} B_L(q) B_R(q)
\frac{1}{2} [ \frac {1} {q^2-m_{ {\tilde q}_1^2}} - 
\frac{1}{q^2-m_{{\tilde q}_2^2}}]
$$
$$
({\bar q}_L \chi_{1R} {\bar \chi}_{1L} q_R + 
{\bar q}_R \chi_{1L} {\bar \chi}_{1R} q_L)
$$
Employing a Fierz transformation we can cast it in the form
\barr
\lefteqn{{\it L}_{eff} = - \frac{G_F} {\sqrt{2}} [\beta_+(q) ({\bar q} q 
{\bar \chi}_1 \chi_1
+ {\bar q}\gamma_5 q {\bar \chi}_1 \gamma_5\chi_1 -({\bar q}\sigma_{\mu\nu} q)
({\bar \chi}_1 \sigma^{\mu\nu}\chi_1))} & & \nonumber \\
 & &+ \beta_- ({\bar q} \tau_3 q {\bar \chi}_1 \chi_1
+ {\bar q} \tau_3 \gamma_5 q {\bar \chi}_1 \gamma_5 \chi_1 - {\bar q} 
\sigma_{\mu\nu} \tau_3 q {\bar \chi}_1 \sigma^{\mu\nu} \chi_1)] \nonumber   
\earr
where for the light quarks u and d
\barr
\beta_{\pm} &=& \frac{1}{3} tan\theta_W C_{11} \{ 2sin 2\theta_{\tilde u}
[\frac{1}{6} C^R_{11} tan\theta_W + \frac{1}{2} C^R_{21}] \Delta_{\tilde u}
 \nonumber \\
 &\mp& sin 2 \theta_{\tilde d}
[\frac{1}{6} C^R_{11} tan\theta_W -  \frac{1}{2} C^R_{21}] \Delta_{\tilde d} \}  
\nonumber
\earr
for quarks other than u and d we only have only the isoscalar
contribution which is given by 
\barr
\beta_{+} &=& \frac{2}{3} tan\theta_W C_{11} \{ 2sin 2\theta_{\tilde u}
[\frac{1}{6} CR_{11} tan\theta_W + \frac{1}{2} CR_{21}] \Delta_{\tilde u}
 \nonumber \\
 &+& sin 2 \theta_{\tilde d}
[\frac{1}{6} CR_{11} tan\theta_W -  \frac{1}{2} C^R_{21}] \Delta_{\tilde d} \}  
\nonumber
\earr
Where in the last expression $u$ indicates quarks with
charge 2/3 and d quarks with charge -1/3. In all cases
\begin{center}
$\Delta_{\tilde u} = \frac{(m^2_{{\tilde u}_1}-m^2_{{\tilde u}_2}) M^2_W}
{(m^2_{{\tilde u}_1}-q^2)(m^2_{{\tilde u}_2}-q^2)}$
\end{center}
and an analogous equation for $\Delta_{\tilde d}$. 

The appearance of scalar terms in s-quark exchange has been first noticed
by Griest.~\cite{ref1} It has also been noticed there that one should consider
explicitly the effects of quarks other than u and d~\cite{Dree} in going 
from the quark to the nucleon level. We first notice that with the exception 
of $t$ s-quark the ${\tilde q}_L - {\tilde q}_R$ mixing small. Thus

\barr
sin 2\theta_{\tilde u} \Delta {\tilde u} &\simeq& \frac{2m_u(A + \mu cot\beta)
m^2_W}  {(m^2_{{\tilde u}_L}-q^2) (m^2_{{\tilde u}_R}-q^2)}\nonumber \\
sin 2\theta_{\tilde d} \Delta {\tilde d} &\simeq& \frac{2m_d(A + \mu tan\beta)
m^2_W}  {(m^2_{{\tilde d}_L}-q^2) (m^2_{{\tilde d}_R}-q^2)}\nonumber
\earr
In going to the nucleon level and ignoring the negligible pseudoscalar and
tensor components we only need modify the above expressions for all
quarks other than t by the substitution $m_q \rightarrow f_q m_N$.
 We will see in the next section that the quarks s,c and b tend to
dominate. For the t s-quark the mixing is complete,
which implies that the amplitude is independent of the top quark mass.
Hence in the case of the top quark we do not get an extra enhancement in
going from the quark to the nucleon level. 
In any case this way we get   
\beq
{\it L}_{eff} = \frac{G_F} {\sqrt {2}} [f^0_s ({\tilde q}) {\bar N} N +
 f^1_s ({\tilde q}) {\bar N} \tau_3 N] {\bar \chi}_1 \chi_1
\eeq
with 
\beq
f^0_s ({\tilde q}) = f^0_q \beta_+  \ \ \  and \ \ \
 f^1_s ({\tilde q}) = f^1_q \beta_- \ \ 
\eeq
(see sect. 3.3 for details).
In the allowed SUSY parameter space considered in this
work this contribution can be neglected in front of the Higgs exchange
contribution. This happens because for quarks other than t the s-quark
mixing is small. For the t-quark, as it has already been mentioned, we
have large mixing, but we do not get the advantage of the mass enhancement.
 
\bigskip
\subsection{The Intermediate Higgs Contribution}
\bigskip

The coherent scattering can be mediated via the intermediate Higgs
particles which survive as physical particles .
The relevant interaction can arise out of the
Higgs-Higgsino-gaugino interaction which takes the form
\barr
 {\it L}_{H \chi \chi} &=& \frac {g}{\sqrt 2} \Big({\bar{\tilde W}}^3_R
 {\tilde H}_{2L} H^{0*}_2 - {\bar{\tilde W}}^3_R {\tilde H}_{1L} H^{0*}_1 
\nonumber \\
   &-&tan \theta_w ({\bar{\tilde B}}_R
  {\tilde H}_{2L} H^{0*}_2 - {\bar{\tilde B}}_R {\tilde H}_{1L} H^{0*}_1)
 \Big) + H.C.
 \label{eq:eg 26}
\earr
Proceeding as above we can express ${\tilde W}$ an ${\tilde B}$ in terms
of the appropriate eigenstates and retain the LSP to obtain
\barr
{\it L} &=& \frac {g}{\sqrt 2} \Big((C^R_{21} -tan\theta_w C^R_{11})
C_{41}{\bar \chi}_{1R} \chi_{1L} H^{o*}_2 \nonumber \\
  &-&(C^R_{21} -tan \theta_w C^R_{11}) 
C_{31}{\bar \chi}_{1R} \chi_{1L} H^{o*}_1 \Big) + H.C.
\label{eq:eg 27}
\earr

We can now proceed further and express the fields 
${H^0_1}^*$, ${H^0_2}^*$ in terms of the physical fields $h$, $H$ and
$A.$ The term which contains $A$ will be neglected, since it yields only
a pseudoscalar coupling which does not lead to coherence.

Thus we can write
\beq
{\cal L}_{eff} = - \frac{G_F}{\sqrt{2}}{\bar \chi} \chi \,
{\bar N} [ f^0_s (H) + f^1_s (H) \tau_3 ] N
\label{2.1.1}
\eeq
where
\beq
f^0_s (H)  = \frac{1}{2} (g_u + g_d) + g_s + g_c + g_b + g_t
\label{2.1.2}
\eeq
\beq
f^1_s (H)  = \frac{1}{2} (g_u - g_d)
\label{2.1.3}
\eeq
with
\beq
g_{a_i}  = \big[ g_1(h) \frac{cos \alpha}{sin \beta}
+ g_2(H) \frac{sin \alpha}{sin \beta} \big] \frac{m_{a_i}}{m_N},
\quad a_i = u,c,t
\label{2.1.4}
\eeq
\beq
g_{\kappa_i}  = \big[- g_1(h) \frac{sin \alpha}{cos \beta}
+ g_2(H) \frac{cos \alpha}{cos \beta} \big] 
\frac{m_{\kappa_i}}{m_N},
\quad \kappa_i = d,s,b
\label{2.1.5}
\eeq
\beq
g_{1}(h)  = 4 (C^R_{11 } tan \theta_W - C^R_{21}) (C_{41 } cos \alpha + 
            C_{31} sin \alpha) \frac{m_N m_W}{m^2_h -q^2}
\label{2.1.6}
\eeq
\beq
g_{2}(H)  = 4 (C^R_{11 } tan \theta_W - C^R_{21}) (C_{41 } sin \alpha - 
            C_{31}cos \alpha) \frac{m_N m_W}{m^2_H -q^2}
\label{2.1.7}
\eeq
where $m_N$ is the nucleon mass, and
the parameters $m_h$, $m_H$ and $\alpha$ depend on the SUSY parameter
space (see Table 1). If one ignores quarks other than
$u$ and $d$ (model A) and uses $m_u =5 MeV=m_d/2$ finds~\cite{Adler}
\beq
f^0_s = 1.86 (g_u + g_d)/2, \quad
f^1_s = 0.49 (g_u - g_d)/2, \quad
\label{2.1.8}
\eeq

\bigskip
\section{Going from the Quark to the Nucleon Level} 
\bigskip
As we have already mentioned, one has to be a bit more 
careful in handling quarks other than $u$ and $d$ since their couplings
are proportional to their mass~\cite{Dree}.
One encounters in the nucleon not only
sea quarks ($u {\bar u}, d {\bar d}$ and $s {\bar s}$) but the heavier
quarks also due to QCD effects ~\cite{Dree00}. 
This way one obtains the scalar Higgs-nucleon
coupling by using  effective quark masses as follows
\begin{center}
$m_u \ra f_u m_N, \ \ m_d \ra f_d m_N. \ \ \  m_s \ra f_s m_N$   
\end{center}
\begin{center}
$m_Q \ra f_Q m_N, \ \ (heavy\ \  quarks \ \ c,b,t)$   
\end{center}
where $m_N$ is the nucleon mass. The isovector contribution is now
negligible. The parameters $f_q,~q=u,d,s$ can be obtained by chirala
symmetry breaking 
terms in relation to phase shift and dispersion analysis.
Following Cheng and Cheng ~\cite{Chen} we obtain
\begin{center}
$ f_u = 0.021, \quad f_d = 0.037, \quad  f_s = 0.140$ 
\quad  \quad  (model B)   
\end{center}
\begin{center}
$ f_u = 0.023, \quad f_d = 0.034, \quad  f_s = 0.400$ 
\quad  \quad  (model C)   
\end{center}
 We see that in both models the s-quark is dominant.
Then to leading order via quark loops and gluon exchange with the
nucleon one finds:
\begin{center}
\quad $f_Q= 2/27(1-\quad \sum_q f_q)$   
\end{center}
This yields:
\begin{center}
\quad $ f_Q = 0.060$ \quad  \quad  (model C)   
\end{center}
\begin{center}
\quad $ f_Q = 0.040$ \quad  \quad  (model C)   
\end{center}
 There is a correction to the above parameters coming from loops
involving s-quarks 
\cite {Dree00}.
The leading contribution can be absorbed into
the definition if the functions $g_1(h)$ and $g_2(H)$ as follows :
\begin{center}
\quad $g_1(h) \rightarrow g_1(h)[1+\frac{1}{8}(2 \frac{m^2_Q}{m^2_W}
-\frac{sin(\alpha+\beta)}{cos^2\theta_W}\frac{sin\beta}{cos\alpha})]$   
\end{center}
\begin{center}
\quad $g_2(H) \rightarrow g_1(h)[1+\frac{1}{8}(2 \frac{m^2_Q}{m^2_W}
+\frac{cos(\alpha+\beta)}{cos^2\theta_W}\frac{sin\beta}{sin\alpha})]$   
\end{center}
for $Q=c$ and $t$ For the b-quark we get:
\begin{center}
\quad $g_1(h) \rightarrow g_1(h)[1+\frac{1}{8}(2 \frac{m^2_b}{m^2_W}
-\frac{sin(\alpha+\beta)}{cos^2\theta_W}\frac{cos\beta}{cos\alpha})]$   
\end{center}
\begin{center}
\quad $g_2(H) \rightarrow g_1(h)[1+\frac{1}{8}(2 \frac{m^2_b}{m^2_W}
-\frac{cos(\alpha+\beta)}{cos^2\theta_W}\frac{cos\beta}{sin\alpha})]$   
\end{center}
In addition to the above effects one has to consider QCD effects. 
These effects renormalize the quark loops as follows \cite {Dree00}:
\begin{center}
\quad $f_{QCD}(q)=\frac{1}{4}\frac{\beta(\alpha_s)}
{1+\gamma_m(\alpha_s)}$
\end{center}
with
\begin{center}
$\beta(\alpha_s)=\frac{\alpha_s}{3 \pi}[1+\frac{19}{4}{\alpha_s}{\pi}]$ ,
$\gamma_m(\alpha_s)=2\frac{\alpha_s}{\pi}$
\end{center}
Thus
\begin{center}
\quad $f_{QCD}(q)=1+\frac{11}{4}\frac{\alpha_s}{\pi}$
\end{center}
The QCD correction associated with the s-quark loops is:
\begin{center}
\quad $f_{QCD}(\tilde{q})=1+\frac{25}{6}\frac{\alpha_s}{\pi}$
\end{center}
 The above corrections depend on Q since $\alpha_s$ must be evaluated
at the scale of $m_Q$. 

 It convenient to introduce the factor$f_{QCD}(\tilde{q})/f_{QCD}(q)$
into the factors $g_1(h)$ and $g_2(H)$ and the factor of $f_{QCD}(q)$
into the the quantities $f_Q$. If, however, one restricts oneself to
the large $tan\beta$ regime, the corrections due to the s-quark loops
is independent of the parameters $\alpha$ and $\beta$ and significant
only for the t-quark.

For large $tan \beta$ we find:
\begin{center}
\quad $f_{c}=0.060 \times 1.068=0.064,
       f_{t}=0.060 \times 2.048=0.123,
       f_{b}=0.060 \times 1.174=0.070$\ quad (model B)
\end{center}
\begin{center}
\quad $f_{c}=0.040 \times 1.068=0.043,
       f_{t}=0.040 \times 2.048=0.082,
       f_{b}=0.040 \times 1.174=0.047$\ quad (model B)
\end{center}
For a more detailed discussion we refer the reader to 
Refs.~\cite{Dree,Dree00}

\bigskip
\section{Summary of the Input Parameters.} 
\bigskip

We have seen that, the vector and axial vector form factors can
arise out of Z-exchange and s-quark exchange.~\cite{JDV}-\cite{ref1}
They have uncertainties in them. Here we consider the three choices in the
allowed parameter space of Kane {\it et al}~\cite{ref3} 
and the eight parameter choices of Castano {\it et al}~\cite{ref3}
These involve universal soft breaking masses at the scale. Non-universal
masses have also recently been employed~\cite{Dree} (see also Arnowitt and Nath
\cite{ref4}
In our choice of the parameters the LSP is mostly a gaugino. Thus, the Z-
contribution is small. It may become dominant in models in which the LSP
happens to be primarily a Higgsino. Such models, however, are excluded by the
cosmological bounds on the relic abundance of LSP.
The transition from the quark to the nucleon level is pretty straightforward 
in the case of vector current contribution. 
We will see later that, due to the Majorana nature of the LSP, the
contribution of the vector current, which can lead to a coherent effect of all
nucleons, is suppressed.~\cite{JDV} 
The vector current is effectively multiplied by a factor of $\beta=v/c$, 
$v$ is the velocity of LSP (see Tables 2, 3 ).
Thus, the axial current, especially in the
case of light and medium mass nuclei, cannot be ignored.

\begin{table}[t]  
\caption{  
The coupling constants entering ${\cal L}_{eff}$, Eqs.
(\ref{eq:eg 41}), (\ref{eq:eg.42}) and (\ref{eq:eg.45}) of the text, 
for solutions \#1 - \#3.
}

 \begin{tabular}{|l|ccc|}
\hline
\hline
 &  &  &  \\
Quantity & Solution $\#1$ & Solution $\#2$ & Solution $\#3$  \\
\hline
 &  &  &  \\
$\beta f^0_V$ &$1.746\times10^{-5}$&$2.617\times10^{-5}$
 & $2.864\times10^{-5}$ \\
$f^1_V/f^0_V$ &-0.153 & -0.113 &-0.251 \\
$f^0_S(H)$ (model A) &$1.31\times10^{-5}$&$1.30\times10^{-4}$ 
   &$1.38\times10^{-5}$ \\
$f^1_S/f^0_S$ (model A) &-0.275 &-0.107 &-0.246 \\
$f^0_S(H)$ (model B) &$5.29\times10^{-4}$&$7.84\times10^{-3}$ 
   &$6.28\times10^{-4}$ \\
$f^0_S(H)$ (model C) &$7.57\times10^{-4}$&$7.44\times10^{-3}$ 
   &$7.94\times10^{-4}$ \\
$f^0_A (NQM) $& $0.510\times10^{-2}$&$3.55\times10^{-2}$
 & $0.704\times10^{-2}$ \\
$f^0_A (EMC) $& $0.612\times10^{-3}$&$0.426\times10^{-2}$
 & $0.844\times10^{-3}$ \\
$f^1_A $& $1.55\times10^{-2}$&$5.31\times10^{-2}$
 & $3.00\times10^{-2}$ \\ 
\hline
\hline
\end{tabular}
\end{table}
\begin{table}[t]  
\caption{  
The same as in Table 2 for solutions \#4 - \#9.
}
\begin{center}
\begin{tabular}{|l|llllll|}
\hline
\hline
  & & & & & &  \\
 Solution   &  $\#$4 & $\#$5  & $\#$6 & $\#$7  & $\#$8  & $\#$9 \\
\hline
  & & & & & &  \\
$\langle \beta^2\rangle^{1/2} f_V^0 $& 0.225 $10^{-4}$ & 0.190 $10^{-4}$&
0.358 $10^{-4}$ & 0.108 $10^{-4}$ & 0.694 $10^{-4}$ & 0.864 $10^{-4}$  \\
$f_V^1/f^0_V$ & -0.0809 & -0.0050 & -0.0320 &
                -0.0538 & -0.0464 & -0.0369   \\
$f_S^0(A)$  &-0.179 $ 10^{-4}$&-0.236 $ 10^{-4}$&
             -0.453 $ 10^{-4}$&-0.266 $ 10^{-4}$&
             -0.210 $ 10^{-3}$&-0.131 $ 10^{-3}$  \\
$f_S^0(B)$  &-0.531 $ 10^{-2}$&-0.145 $ 10^{-2}$&
             -0.281 $ 10^{-2}$&-0.132 $ 10^{-1}$&
             -0.117 $ 10^{-1}$&-0.490 $ 10^{-2}$  \\
$f_S^0(C)$  &-0.315 $ 10^{-2}$&-0.134 $ 10^{-2}$&
             -0.261 $ 10^{-2}$&-0.153 $ 10^{-1}$&
             -0.118 $ 10^{-1}$&-0.159 $ 10^{-2}$  \\
$f_S^1(A)$  &-0.207 $ 10^{-5}$&-0.407 $ 10^{-5}$&
              0.116 $ 10^{-4}$& 0.550 $ 10^{-4}$&
              0.307 $ 10^{-4}$& 0.365 $ 10^{-4}$  \\
$f_A^0(NQM)$& 6.950 $ 10^{-3}$& 5.800 $ 10^{-3}$&
              1.220 $ 10^{-2}$& 3.760 $ 10^{-2}$&
              3.410 $ 10^{-2}$& 2.360 $ 10^{-2}$  \\
$f_A^0(EMC)$& 0.834 $ 10^{-3}$& 0.696 $ 10^{-3}$&
              0.146 $ 10^{-2}$& 0.451 $ 10^{-2}$&
              0.409 $ 10^{-2}$& 0.283 $ 10^{-2}$  \\
$f_A^1$     & 2.490 $ 10^{-2}$& 1.700 $ 10^{-2}$&
              3.440 $ 10^{-2}$& 2.790 $ 10^{-1}$&
              1.800 $ 10^{-1}$& 2.100 $ 10^{-1}$  \\
\hline
\hline
\end{tabular}
\end{center}
\end{table}
For the isovector axial current one is pretty confident about how to go 
from the quark to the nucleon level. We know from ordinary weak decays
that the coupling merely gets renormalized from $g_A=1$ to $g_A=1.24$.
For the isoscalar axial current the situation is not completely clear.
The naive quark model (NQM) would give a renormalization parameter of unity
(the same as the isovector vector current). This point of view has, 
however, changed in recent years due to the so-called spin
crisis,~\cite{Ashm}-\cite{Gensin} i.e.
the fact that in the EMC data~\cite{Ashm} it appears that only a small fraction
of the proton spin arises from the quarks. Thus, one may have to
renormalize $f^0_A$ by $g^0_A=0.28$, for u and d quarks, and $g^0_A=-0.16$
for the strange quarks,~\cite{Jaffe,Gensin} i.e. a total factor of 0.12. 
These two possibilities, labeled as NQM and EMC, are listed in 
Tables 2,3. One cannot completely rule out the possibility that the 
actual value maybe anywhere in the above mentioned region.~\cite{Gensin}

The scalar form factors arise out of the Higgs exchange or via s-quark 
exchange when there is mixing~\cite{Dree} between s-quarks ${\tilde q}_L$
and ${\tilde q}_R$
(the partners of the left-handed and right-handed quarks).
We have seen ~\cite{JDV} that they have two types 
of uncertainties in them. One, which is the most important,
at the quark level due to the uncertainties in the Higgs sector. 
The actual values of the parameters $f^0_S$ and $f^1_S$ used here, arising 
mainly from Higgs exchange, were obtained by considering 1-loop corrections
in the Higgs sector. As a result, the lightest Higgs mass is now a bit higher,
i.e. more massive than the value of the Z-boson.~\cite{El-Ri,Haber}

The other type of uncertainty is related to the step going from  
the quark to the nucleon level~\cite{Dree} (see sect. 3.3).
Such couplings are proportional to the
quark masses, and hence sensitive to the small admixtures of $q {\bar q }$
(q other than u and d) present in the nucleon. Again values of $f^0_S$ and
$f^1_S$ in the allowed SUSY parameter space are considered (see Tables
 2,3).

\section{Expressions for the Unconvoluted Event Rates.} 
\bigskip

Combining for results of the previous section we can write
 \beq
{\it L}_{eff} = - \frac {G_F}{\sqrt 2} \{({\bar \chi}_1 \gamma^{\lambda}
\gamma_5 \chi_1) J_{\lambda} + ({\bar \chi}_1 
 \chi_1) J\}
 \label{eq:eg 41}
\eeq
where
\beq
  J_{\lambda} =  {\bar N} \gamma_{\lambda} (f^0_V +f^1_V \tau_3
+ f^0_A\gamma_5 + f^1_A\gamma_5 \tau_3)N
 \label{eq:eg.42}
\eeq
with
\barr
 f^0_V  &=&  f^0_V(Z) + f^0_V({\tilde q}),  \qquad
f^1_V  =  f^1_V(Z) + f^1_V({\tilde q})
\nonumber \\
  f^0_A  &=&  f^0_A(Z) + f^0_A({\tilde q}) , \qquad
f^1_A  =  f^1_A(Z) + f^1_A({\tilde q})
 \label{eq:eg 43}
\earr
and
\beq
J = {\bar N} (f^0_s +f^1_s \tau_3) N
 \label{eq:eg.45}
\eeq

We have neglected the uninteresting pseudoscalar and tensor
currents. Note that, due to the Majorana nature of the LSP, 
${\bar \chi_1} \gamma^{\lambda} \chi_1 =0$ (identically).

 With the above ingredients the differential cross section can be cast in the 
form 
\begin{equation}
d\sigma (u,\upsilon)= \frac{du}{2 (\mu _r b\upsilon )^2} [(\bar{\Sigma} _{S} 
                   +\bar{\Sigma} _{V}~ \frac{\upsilon^2}{c^2})~F^2(u)
                       +\bar{\Sigma} _{spin} F_{11}(u)]
\label{2.9}
\end{equation}
\begin{equation}
\bar{\Sigma} _{S} = \sigma_0 (\frac{\mu_r}{m_N})^2  \,
 \{ A^2 \, [ (f^0_S - f^1_S \frac{A-2 Z}{A})^2 \, ] \simeq \sigma^S_{p,\chi^0}
        A^2 (\frac{\mu_r}{m_N})^2 
\label{2.10}
\end{equation}
\begin{equation}
\bar{\Sigma} _{spin}  =  \sigma^{spin}_{p,\chi^0}~\zeta_{spin}
\label{2.10a}
\end{equation}
\begin{equation}
\zeta_{spin}= \frac{(\mu_r/m_N)^2}{3(1+\frac{f^0_A}{f^1_A})^2}
[(\frac{f^0_A}{f^1_A} \Omega_0(0))^2 \frac{F_{00}(u)}{F_{11}(u)}
  +  2\frac{f^0_A}{ f^1_A} \Omega_0(0) \Omega_1(0)
\frac{F_{01}(u)}{F_{11}(u)}+  \Omega_1(0))^2  \, ] 
\label{2.10b}
\end{equation}
\begin{equation}
\bar{\Sigma} _{V}  =  \sigma^V_{p,\chi^0}~\zeta_V 
\label{2.10c}
\end{equation}
\begin{equation}
\zeta_V = \frac{(\mu_r/m_N)^2} {(1+\frac{f^1_V}{f^0_V})^2} A^2 \, 
(1-\frac{f^1_V}{f^0_V}~\frac{A-2 Z}{A})^2 [ (\frac{\upsilon_0} {c})^2  
[ 1  -\frac{1}{(2 \mu _r b)^2} \frac{2\eta +1}{(1+\eta)^2} 
\frac{\langle~2u~ \rangle}{\langle~\upsilon ^2~\rangle}] 
\label{2.10d}
\end{equation}
\\
$\sigma^i_{p,\chi^0}=$ proton cross-section,$i=S,spin,V$ given by:\\
$\sigma^S_{p,\chi^0}= \sigma_0 ~(f^0_S)^2$   (scalar) , 
(the isovector scalar is negligible, i.e. $\sigma_p^S=\sigma_n^S)$\\
$\sigma^{spin}_{p,\chi^0}= \sigma_0~~3~(f^0_A+f^1_A)^2$
  (spin) ,
$\sigma^{V}_{p,\chi^0}= \sigma_0~(f^0_V+f^1_V)^2$  
(vector)   \\
where $m_p$ is the proton mass,
 $\eta = m_x/m_N A$, and
 $\mu_r$ is the reduced mass and  
\begin{equation}
\sigma_0 = \frac{1}{2\pi} (G_F m_N)^2 \simeq 0.77 \times 10^{-38}cm^2 
\label{2.7} 
\end{equation}
\begin{equation}
u = q^2b^2/2
\label{2.15} 
\end{equation}
or equivalently
\begin{equation}
Q=Q_{0}u, \qquad Q_{0} = \frac{1}{A m_{N} b^2} 
\label{2.16b} 
\end{equation}
where
b is (the harmonic oscillator) size parameter, 
q is the momentum transfer to the nucleus, and
Q is the energy transfer to the nucleus (see Table 4)\\
In the above expressions $F(u)$ is the nuclear form factor and
\begin{equation}
F_{\rho \rho^{\prime}}(u) =  \sum_{\lambda,\kappa}
\frac{\Omega^{(\lambda,\kappa)}_\rho( u)}{\Omega_\rho (0)} \,
\frac{\Omega^{(\lambda,\kappa)}_{\rho^{\prime}}( u)}
{\Omega_{\rho^{\prime}}(0)} 
, \qquad \rho, \rho^{\prime} = 0,1
\label{2.11} 
\end{equation}
are the spin form factors \cite{KVprd} ($\rho , \rho^{'}$ are isospin indices)\\
Both form factors are normalized to one at $u=0$.\\
$\Omega_0$ ($\Omega_1$) are the static isoscalar (isovector) spin 
matrix elements (see tables 5 and 6).

\begin{table}[t]  
\caption{  
 The quantity $q_0$ (forward momentum transfer) in units of
$fm^{-1}$ for three values of $m_{\chi}$ and three typical nuclei.
In determining $q_0$ the value $\langle \beta^2 \rangle^{1/2} =10^{-3}$ was
employed.
}
\begin{tabular}{|c|ccc|}
\hline
\hline
  &  &  &  \\
& \multicolumn{3}{|c|}{ $q_0$ ($fm^{-1}$)}  \\
\hline
  &  &  &  \\
Nucleus & $m_{\chi}=30.0 \, GeV$ & $m_{\chi}=100.0 \, GeV$ & $m_{\chi}=150.0 \, GeV$ \\
\hline
  &  &  &  \\
  $Ca$  &  0.174   &  0.290 &  0.321 \\
  $Ge$  &  0.215   &  0.425 &  0.494 \\
  $Pb$  &  0.267   &  0.685 &  0.885 \\ 
\hline
\hline
\end{tabular}
\end{table}
\begin{table}[t]  
\caption{  
Comparison of the static spin matrix elements for
three typical nuclei, $Pb$ (present  calculation) and $^{73}Ge$, 
$^{19}F$ , $^{23}Na$, $^{29}Si $
(see Ref. \cite{Ress,DIVA00}).
}
\begin{center}
\begin{tabular}{|l|rrrrr|}
\hline
\hline
  &  &  &  & & \\
Component & $^{207}Pb_{{1/2}^-}$ & $^{73}Ge_{{9/2}^+}$ 
 &$^{19}Si_{{1/2}^+}$ &$ ^{23}Na_{{3/2}^+}$&$^{29}Si_{{1/2}^+}$ \\
\hline
  &  &  &  & & \\
$\Omega^2_1(0)$ & 0.231 & 1.005 & 2.807 & 0.346&0.220 \\
$\Omega_1(0) \Omega_0(0)$ & -0.266 & -1.078 & 2.707 & 0.406 & -0.214 \\
$\Omega^2_0(0)$ & 0.305 & 1.157 &   2.610 & 0.478 &0.208 \\
\hline
\hline
\end{tabular}
\end{center}
\end{table}
\begin{table}[t]  
\caption{  
Ratio of spin contribution ($^{207}Pb/^{73}Ge$) at the relevant
momentum transfer with the kinematical factor $1/(1+\eta)^2, \,\, 
\eta=m/A m_N.$
}
\begin{center}
\begin{tabular}{|c|lllllllll|}
\hline
\hline
 & & & & & & & &  &   \\
 Solution & $\#$1 & $\#$2 & $\#$3 & $\#$4 & $\#$5 & $\#$6 & $\#$7 & $\#$8 
 & $\#$9 \\
\hline
 & & & & & & & &  &   \\
$m_x \,(GeV)$ & 126 & 27 & 102 & 80 & 124 & 58 & 34 & 35 & 50 \\
\hline
 & & & & & & & &  &   \\
NQM & 0.834 & 0.335 & 0.589 & 0.394 & 0.537 & 0.365 & 0.346 & 0.337 & 0.417 \\
EMC & 0.645 & 0.345 & 0.602 & 0.499 & 0.602 & 0.263 & 0.341 & 0.383 & 0.479 \\
\hline
\hline
\end{tabular}
\end{center}
\end{table}
 The non-directional event rate is given by:
\begin{equation}
R=R_{non-dir} =\frac{dN}{dt} =\frac{\rho (0)}{m_{\chi}} \frac{m}{A m_N} 
\sigma (u,\upsilon) | {\boldmath \upsilon}|
\label{2.17} 
\end{equation}
 Where
 $\rho (0) = 0.3 GeV/cm^3$ is the LSP density in our vicinity and 
 m is the detector mass 
 
The differential non-directional  rate can be written as
\begin{equation}
dR=dR_{non-dir} = \frac{\rho (0)}{m_{\chi}} \frac{m}{A m_N} 
d\sigma (u,\upsilon) | {\boldmath \upsilon}|
\label{2.18}  
\end{equation}
where $d\sigma(u,\upsilon )$ was given above.

 The directional differential rate \cite{Copi99} in the direction
$\hat{e}$ is 
given by :
\begin{equation}
dR_{dir} = \frac{\rho (0)}{m_{\chi}} \frac{m}{A m_N} 
{\boldmath \upsilon}.\hat{e} H({\boldmath \upsilon}.\hat{e})
 ~\frac{1}{2 \pi}~  
d\sigma (u,\upsilon)
\label{2.20}  
\end{equation}
where H the Heaviside step function. The factor of $1/2 \pi$ is 
introduced, since  the differential cross section of the last equation
is the same with that entering the non-directional rate, i.e. after
an integration
over the azimuthal angle around the nuclear momentum has been performed.
In other words, crudely speaking, $1/(2 \pi)$ is the suppression factor we
 expect in the directional rate compared to the usual one. The precise 
suppression factor depends, of course, on the direction of observation.
 
\section{Convolution of the Event Rates.} 
\bigskip
 
 We have seen that the event rate for LSP-nucleus scattering depends on the
relative LSP-target velocity. In this section we will examine the consequences 
of the earth's
revolution around the sun (the effect of its rotation around its axis is
expected to
be negligible) i.e. the modulation effect. In practice this has been 
accomplished by
assuming a consistent LSP velocity dispersion, such as  
a Maxwell distribution ~\cite{Jungm}. More recently other non-isothermal
approaches, in the context velocity peaks and caustic rings, have been 
proposed, see e.g Sikivie
et al \cite{SIKIVIE}. 
 Let us begin with isothermal models. In the present paper following
the work of Drukier, see Ref. ~\cite{Druk},
 we will assume that the velocity distribution is
only axially symmetric, i.e. of the form
\beq
f(\upsilon^{\prime},\lambda) =  N(y_{esc},\lambda)( \sqrt{\pi}\upsilon_0)^{-3}) 
                              [ f_1(\upsilon^{\prime},\lambda)-
                                f_2(\upsilon^{\prime},\upsilon_{esc},\lambda)]
\label{3.3}  
\eeq
with
\beq
f_1(\upsilon^{\prime},\lambda)=exp[(- \frac{(\upsilon^{\prime}_x)^2+
                             (1+\lambda)((\upsilon^{\prime}_y)^2 +
                              (\upsilon^{\prime}_z)^2)}{\upsilon^2_0}]
\label{3.4}  
\eeq
\beq
f_2(\upsilon^{\prime},\upsilon_{esc},\lambda)=exp[- \frac{\upsilon^2_{esc}+
                                             \lambda((\upsilon^{\prime}_y)^2 +
                              (\upsilon^{\prime}_z)^2)}{\upsilon^2_0}]
\label{3.5}  
\eeq
where
\beq
v_0 = \sqrt{(2/3) \langle v^2 \rangle } =220 Km /s
\label{3.2}  
\eeq
i.e. ${ v}_0$ is the velocity of the sun around the center of the galaxy.
 $\upsilon_{esc}$ is the escape velocity in the
gravitational field of the galaxy, $\upsilon_{esc}=625Km/s$ \cite {Druk}.
In the above expressions $\lambda$ is a parameter, which describes the 
asymmetry and takes values between 0 and 1 and  N is a proper normalization 
constant \cite{JDV99b}.
For $y_{esc} \rightarrow \infty$ we get the simple expression 
$N^{-1}=\lambda+1$

 In a recently proposed non-isothermal model one consider the late 
 in-fall of dark matter
into our  galaxy producing flows of caustic rings. In particular the predictions of a self-similar model have been put forward as a possible scenario
for dark matter density-velocity distribution, see e.g. Sikivie et al
\cite{SIKIVIE}. The implications of such theoretical predictions and,
in particular, the modulation effect are the subject of this section. 

 Following Sikivie we will consider $2 \times N$ caustic rings, (i,n)
, i=(+.-) and n=1,2,...N (N=20 in the model of Sikivie et al),
each of which
contributes to the local density a fraction $\bar{\rho}_n$ of the
of the summed density $\bar{\rho}$ of each of the $i=+,-$. It contains
WIMP like particles with velocity 
${\bf y}^{'}_n=(y^{'}_{nx},y^{'}_{ny},y^{'}_{nz})$
in units of essentially
the sun's velocity ($\upsilon_0=220~Km/s$), with respect to the
galactic center.
The z-axis is chosen in the direction of the disc's rotation,
 i.e. in the direction of the motion of the
the sun, the y-axis is perpendicular to the plane of the galaxy and 
the x-axis is in the radial direction. We caution the reader that
these axes are traditionally indicated by astronomers as
$\hat{e}_{\phi},\hat{e}_r, \hat{e}_z)$ respectively. 
The needed quantities are taken from the 
work of Sikivie et al \cite{SIKIVIE} (see Table 7), via the 
definitions

$y^{'}_n=\upsilon_n/\upsilon_0
,y^{'}_{nz}=\upsilon_{n\phi}/\upsilon_0=y_{nz}
,y^{'}_{nx}=\upsilon_{nr}/\upsilon_0=y_{nx}
,y^{'}_{ny}=\upsilon_{nz}/\upsilon_0=y_{ny}
,\rho_{n}=d_n/\bar{\rho}
,\bar{\rho}=\sum_{n=1}^N~d_n$ and 
$y_n=[(y_{nz}-1)^2+y^2_{ny}+y^2_{nx}]^{1/2}$ (for each flow +.-).
\begin{table}[t]  
\caption{The velocity parameters  
$y^{'}_n =\upsilon_n/\upsilon_0,
~y_{nz} =y^{'}_{nz}=\upsilon_{n \phi}/\upsilon_0,
~y_{ny} =y^{'}_{ny}=\upsilon_{nz}/\upsilon_0,
~y_{nx} =y^{'}_{nx}=\upsilon_{nr}/\upsilon_0$ and
$y_n=[(y_{nz}-1)^2+y_{ny}^2+y_{nx}^2]^{1/2}$. Also shown are the quantities:
$a_n$, the caustic rind radii, and $\bar {\rho}_n=d_n/[\sum _{n=1}^{20} d_n]$. 
( For the other definitions see text ). 
}
\begin{center}
\footnotesize
\begin{tabular}{|l|c|rrrrrr|}
\hline
\hline
& & & & & &      \\
n &  $a_{n}(Kpc)$  & $y^{'}_n$  & $y_{nz}$ & $y_{ny}$  & $y_{nx}$ 
 & $y_n$ & $\bar {\rho}_n $\\
\hline 
& & & & & & &      \\
1 &38.0& 2.818& 0.636& $\pm$2.750& 0.000& 2.773 & 0.0120\\
2 &19.0& 2.568& 1.159& $\pm$2.295& 0.000& 2.301 & 0.0301\\
3 &13.0& 2.409& 1.591& $\pm$1.773& 0.000& 1.869 & 0.0601\\
4 & 9.7& 2.273& 2.000& $\pm$1.091& 0.000& 1.480 & 0.1895\\
5 & 7.8& 2.182& 2.000& 0.000& $\pm$0.863& 1.321 & 0.2767\\
6 & 6.5& 2.091& 1.614& 0.000& $\pm$1.341& 1.475 & 0.0872\\
7 & 5.6& 2.023& 1.318& 0.000& $\pm$1.500& 1.533 & 0.0571\\
8 & 4.9& 1.955& 1.136& 0.000& $\pm$1.591& 1.597 & 0.0421\\
9 & 4.4& 1.886& 0.977& 0.000& $\pm$1.614& 1.614 & 0.0331\\
10& 4.0& 1.818& 0.864& 0.000& $\pm$1.614& 1.619 & 0.0300\\
11& 3.6& 1.723& 0.773& 0.000& $\pm$1.614& 1.630 & 0.0271\\
12& 3.3& 1.723& 0.682& 0.000& $\pm$1.591& 1.622 & 0.0241\\
13& 3.1& 1.619& 0.614& 0.000& $\pm$1.568& 1.615 & 0.0211\\
14& 2.9& 1.636& 0.545& 0.000& $\pm$1.545& 1.611 & 0.0180\\
15& 2.7& 1.591& 0.500& 0.000& $\pm$1.500& 1.581 & 0.0180\\
16& 2.5& 1.545& 0.454& 0.000& $\pm$1.477& 1.575 & 0.0165\\
17& 2.4& 1.500& 0.409& 0.000& $\pm$1.454& 1.570 & 0.0150\\
18& 2.2& 1.455& 0.386& 0.000& $\pm$1.409& 1.537 & 0.0150\\
19& 2.1& 1.432& 0.364& 0.000& $\pm$1.386& 1.525 & 0.0135\\
20& 2.0& 1.409& 0.341& 0.000& $\pm$1.364& 1.515 & 0.0135\\
\hline
\hline
\end{tabular}
\end{center}
\end{table}
 This leads to a
velocity distribution of the form:
\beq
f(\upsilon^{\prime}) = \sum_{n=1}^N~\delta({\bf \upsilon} ^{'}
    -\upsilon_0~{\bf y}^{'}_n)
\label{3.1}  
\eeq
 The actual situation, of course, could be a combination of an
isothermal contribution and late infall of dark matter \cite{Copi99}.
In the present treatment we will consider each of these distributions
separately. For clarity of presentation we will consider each case 
separately.

Since the axis of the ecliptic \cite{KVprd}. 
lies very close to the $y,z$ plane the velocity of the earth around the
sun is given by 
\beq
{\bf \upsilon}_E \, = \, {\bf \upsilon}_0 \, + \, {\bf \upsilon}_1 \, 
= \, {\bf \upsilon}_0 + \upsilon_1(\, sin{\alpha} \, {\bf \hat x}
-cos {\alpha} \, cos{\gamma} \, {\bf \hat y}
+cos {\alpha} \, sin{\gamma} \, {\bf \hat z} \,)
\label{3.6}  
\eeq
where $\alpha$ is the phase of the earth's orbital motion, $\alpha =2\pi 
(t-t_1)/T_E$, where $t_1$ is around second of June and $T_E =1 year$.

One can now express the above distribution in the laboratory frame 
\cite{JDV99b}
by writing $ {\bf \upsilon}^{'}={\bf \upsilon} \, + \, {\bf \upsilon}_E \,$ 

\section{Expressions for the Non-directional Differential Event Rate}

The mean value of the non-directional event rate of Eq. (\ref {2.18}), 
is given by
\beq
\Big<\frac{dR}{du}\Big> =\frac{\rho (0)}{m_{\chi}} 
\frac{m}{A m_N}  
\int f({\bf \upsilon}, {\boldmath \upsilon}_E) 
          | {\boldmath \upsilon}|
                       \frac{d\sigma (u,\upsilon )}{du} d^3 {\boldmath \upsilon} 
\label{3.10} 
\eeq
 The above expression can be more conveniently written as
\beq
\Big<\frac{dR}{du}\Big> =\frac{\rho (0)}{m_{\chi}} \frac{m}{Am_N} \sqrt{\langle
\upsilon^2\rangle } {\langle \frac{d\Sigma}{du}\rangle } 
\label{3.11}  
\eeq
where
\beq
\langle \frac{d\Sigma}{du}\rangle =\int
           \frac{   |{\boldmath \upsilon}|}
{\sqrt{ \langle \upsilon^2 \rangle}} f({\boldmath \upsilon}, 
         {\boldmath \upsilon}_E)
                       \frac{d\sigma (u,\upsilon )}{du} d^3 {\boldmath \upsilon}
\label{3.12}  
\eeq

\subsection{No velocity Dispersion-The Case of Caustic Rings}

 In the case of caustic rings the last expression takes the form
\beq
\langle \frac{d\Sigma}{du} \rangle  = \frac{2 \bar{\rho}}{\rho(0)}~
             a^2 [\bar {\Sigma} _{S} \bar {F}_0(u) +
        \frac{\langle \upsilon ^2 \rangle}{c^2}\bar {\Sigma} _{V} \bar {F}_1(u) 
                          +\bar {\Sigma} _{spin} \bar {F}_{spin}(u)]
\label{3.23}  
\eeq
We remind the reader that $\bar{\rho}$ was obtain for each type of flow
(+ or -), which explains  the factor of two. In the Sikivie model
\cite {SIKIVIE} we have $(2\bar{\rho}/\rho(0)=1.25$, i.e. the whole dark
matter density lies in the form of caustic rings. In a composite model
this can only be a fraction of the total density.

 The quantities
 $\bar{\Sigma} _{i},i=S,V,spin$ are given by Eqs. (\ref {2.10})-
(\ref {2.10c}). 
The quantities 
$\bar{F}_0,\bar{F}_1,\bar{F}_{spin}$  
are obtained from the corresponding form factors via the equations
\beq
\bar{F}_{k}(u) = F^2(u)\bar{\Psi}_k(u) \frac{(1+k) }{2k+1}~~,~~ k = 0,1
\label{3.24}  
\eeq
\beq
\bar{F}_{spin}(u) = F_{11}(u)\bar{\Psi}_0(u) 
\label{3.24a}  
\eeq
The functions $\tilde{\Psi}_k(u)$ depend on the model. 
Introducing the parameter
\beq
\delta = \frac{2 \upsilon_1 }{\upsilon_0}\, = \, 0.27,
\label{3.13}
\eeq
in the Sikivie model we find
\barr
\tilde{\Psi}_k(u)& = &\sqrt{\frac{2}{3}}~\sum_{n=1}^N~ \bar{\rho}_n
                   y_n^{2(k-1)}\Theta(\frac{y_n^2}{a^2}-u)
                   [(y_{nz}-1-\frac{\delta}{2}~sin\gamma~cos\alpha)^2
\nonumber\\
                 &+& (y_{ny}+\frac{\delta}{2}~cos\gamma~cos\alpha)^2
                  + (y_{nx}-\frac{\delta}{2}~sin\alpha)^2]^{1/2}
\label{3.26a}  
\earr
with
\beq
a = \frac{1}{\sqrt{2} \mu _rb\upsilon _0}  
\label{3.27}  
\eeq

Combining the above results the non-directional differential rate takes the form
\beq
\langle \frac{dR}{du} \rangle  = \bar{R} \frac{2 \bar{\rho}}{\rho(0)}
                   t~T(u) [1 - \cos \alpha~ H(u) ]
\label{3.31a}  
\eeq
In the above expressions $\bar{R}$ is the rate obtained in the conventional 
approach \cite {JDV} by neglecting the folding with the LSP velocity and the
momentum transfer dependence of the differential cross section, i.e. by
\beq
\bar{R} =\frac{\rho (0)}{m_{\chi}} \frac{m}{Am_N} \sqrt{\langle
v^2\rangle } [\bar{\Sigma}_{S}+ \bar{\Sigma} _{spin} + 
\frac{\langle \upsilon ^2 \rangle}{c^2} \bar{\Sigma} _{V}]
\label{3.39b}  
\eeq
where $\bar{\Sigma} _{i}, i=S,V,spin$ have been defined above, see Eqs
 (\ref {2.10}) - (\ref {2.10c}). 

The factor 
$T(u)$ takes care of the u-dependence of the unmodulated differential rate. It
is defined so that
\beq
 \int_{u_{min}}^{u_{max}} du T(u)=1.
\label{3.30a}  
\eeq
i.e. it is the relative differential rate. $u_{min}$ is determined by 
the energy cutoff due to the performance of the detector,i.e
\beq
u_{min}= \frac{Q_{min}}{Q_0}
\label{3.30c}  
\eeq
while $u_{max}$ is determined the via the relations:
\beq
u_{max}=min(\frac{y^2_{esc}}{a^2},max(\frac{y_{n} ^2}{a^2})~,~ n=1,2,...,N)
\label{3.30b}  
\eeq
On the other hand
$H(u)$ gives the energy transfer dependent
modulation amplitude (relative to the unmodulated amplitude).
The quantity $t$ takes care of the modification of the total rate due to the
nuclear form factor and the folding with the LSP velocity distribution.
 Since the functions $\bar{F}_0(u),\bar{F}_1$ and $\bar{F}_{spin}$ have 
a different dependence on u, the functions $T(u)$and $H(u)$ 
and $t$, in principle, depend somewhat
on the SUSY parameters. If, however, we ignore the small vector 
contribution and assume (i) the scalar and axial (spin) dependence on u is the
same, as seems to be the case for light systems \cite{DIVA00}$^,$\cite{Verg98},
 or (ii) only one
 mechanism (S, V, spin) dominates, the parameter $\bar{R}$ 
contains the dependence on all SUSY parameters. The parameters $t$ and
$T(u)$ depend on  the LSP mass and the nuclear parameters, while the 
$H(u)$ depends only on the parameter $a$. 

\subsection{Velocity Dispersion-Isothermal Models}
Expanding in powers of $ \delta $ , see Eq. (\ref {3.13}) and keeping terms up
 to linear in it we can
manage to perform the angular integrations \cite{JDV99b} in Eq. 
(\ref {3.12})
 and get 
Eq. (\ref {3.23}).
Now the quantities $\tilde{\Psi}_k(u)$ are given by 
\beq
\tilde{\Psi}_k(u) =  [\tilde {\psi} _{(0),k}(a\sqrt{u})+
            0.135 \cos \alpha \tilde{\psi} _{(1),k}(a\sqrt{u})]  
\label{3.26}  
\eeq
and
\beq
\tilde{\psi} _{(l),k}(x)= N(y_{esc},\lambda) e^{-\lambda}(e^{-1} 
   \tilde{\Phi}_{(l),k}(x)-
                  exp[-y^2_{esc}] \tilde{\Phi}^{'}_{(l),k}(x))  
\label{3.28}  
\eeq
\beq
\tilde {\Phi}_{(l),k}(x) =  \frac{2}{\sqrt{6 \pi}} \int_x^{y_{esc}} dy y^{2k-1}
              exp {(-(1+\lambda) y^2)})( \tilde{F}_{l}(\lambda,(\lambda+1) 2 y )
+ \tilde{G}_{l}(\lambda,y)))
\label{3.29}  
\eeq
\beq
\tilde{\Phi}^{'}_{(l)_,k}(x) =  \frac{2}{\sqrt{6 \pi}} \int_x^{y_{esc}} dy y^{2k-1}
                     exp {(-\lambda y^2)})\tilde {G}^{'}_{l}(\lambda,y))
\label{3.30}  
\eeq
In the above expressions 
\beq
\tilde{G}_0(0,y) =0~~~~~ ,~~~~~\tilde{G}_1(0,y)=0
\label { 3.20c}  
\eeq
\beq
\tilde{F}_0(\lambda,x) = (\lambda+1)^{-2}~x~sinh(x)  
\label{3.21a}  
\eeq
\beq
\tilde{F}_1(\lambda,x)  =   (1+\lambda)^{-2}~
   [(2+\lambda)(x/2)+1)~[x~cosh(x)- (2 \lambda+3)~sinh(x)]
\label{3.21b}  
\eeq
note that here $x=(\lambda+1)2y$. 
The functions $\tilde{G}$ cannot be obtained analytically, but they can easily 
be expressed as a rapidly convergent series in $y=\frac{
\upsilon}{\upsilon_0}$, which will not be given here.  

The functions $\tilde{G}^{'}_i(\lambda,y)$, associated with the small second
term of the velocity distribution are obtained similarly \cite{JDV99b}.

The non-directional differential rate takes the form
\beq
\langle \frac{dR}{du} \rangle  =
 \bar{R}\frac{\rho^{'}(0)}{\rho(0)}  t T(u) [(1 + \cos \alpha H(u))] 
\label{3.31b}  
\eeq
With $\bar{R}$ given by Eq. (\ref {3.39b}) and $\rho^{'}(0)$ is the part
of the total LSP density attributed to this mode. 
Note the difference of sign in the definition of the modulation amplitude H 
compared to Eq. (\ref {3.31a}). 

 Here $u_{min}$ is determined by the energy 
cutoff due to the performance of the detector. $u_{max}$ is determined by the 
escape velocity $\upsilon_{esc}$ via the relation:
\beq
u_{max}=\frac{y_{esc} ^2}{a^2}~~ 
\label{3.30d}  
\eeq
 Considering only the scalar interaction we get
 $\bar{R} \rightarrow \bar{R}_{S}$ and
\beq
t~ T(u) =  a^2 F^2(u) \tilde{\psi} _{(0),0}(a \sqrt{u})    
\label{3.48}  
\eeq

For the spin interaction we get a similar expression except that
$\bar{R}\rightarrow \bar{R}_{spin}$ and 
$F^2 \rightarrow F_{11}$.
Finally for completeness we will consider the less important vector
contribution.
 We get $\bar{R} \rightarrow \bar{R}_{V}$ and
\beq
t~ T(u) =  F^2(u) [ \tilde{\psi} _{(0),1}(a \sqrt{u})
                       -\frac{1}{(2 \mu _r b)^2} \frac{2\eta +1}{(1+\eta)^2}
                        u\, \tilde{\psi} _{(0),0}(a \sqrt{u})] \frac{2a^2}{3}    
\label{3.52}  
\eeq
 The quantity $T(u)$ depends on the nucleus through the nuclear form factor
or the spin response function and the parameter $a$. The modulation amplitude
takes the form
\beq
H(u) =0.135\frac{\tilde{\psi}_{(1),k}(a \sqrt{u})}
                {\tilde{\psi} _{(0),k}(a \sqrt{u})}  
\label{3.45}  
\eeq 
Thus in this case the $H(u)$ depends only on $a\sqrt{u}$, which coincides with 
the parameter x of Ref. \cite{Smith1}, i.e.
only on the momentum transfer, the reduced mass and the size of the nucleus.

 Returning to the differential rate it is sometimes convenient to use the
quantity $T(u) H(u)$ rather than $H$, since H(u) may appear artificially 
increasing function of u
due to the faster decrease of $T(u)$ (H(u) was obtained after division by T(u)) 


\section{Expressions for the Directional Differential Event Rate}

 There are now experiments under way aiming at measuring directional 
rates \cite {UKDMC} using TPC counters which permit the observation
of the recoiling nucleus is observed in a certain direction.
From a theoretical point of view the directional rates have been
previously discussed by Spegel \cite {Sperg96} and Copi {\ et al}
\cite{Copi99}. 
The rate will depend on the direction of observation, showing a strong
correlation with the direction of the sun's motion. In a favorable 
situation the rate will merely be suppressed by a factor of $2 \pi$
relative to the non-directional rate. This is due to the fact that one 
does not now integrate over the
azimuthal angle of the nuclear recoiling momentum. The directional rate
will also show modulation due to the Earth's motion. We will again 
examine an non-isothermal non symmetric case ( the Sikivie model) and
a 3-dimensional Gaussian distribution with only axial symmetry.

The mean value of the directional differential event rate of Eq. (\ref {2.20}), 
is defined by
\beq
\Big<\frac{dR}{du}\Big>_{dir} =\frac{\rho (0)}{m_{\chi}} 
\frac{m}{A m_N} \frac{1}{2 \pi} 
\int f({\bf \upsilon}, {\boldmath \upsilon}_E)
{\boldmath \upsilon}.\hat{e} H({\boldmath \upsilon}.\hat{e})
                       \frac{d\sigma (u,\upsilon )}{du} d^3 {\boldmath \upsilon} 
\label{4.10} 
\eeq
where ${\bf \hat e}$ is the unit vector in the direction of observation. 
It can be more conveniently expressed as
\beq
\Big<\frac{dR}{du}\Big>_{dir} =\frac{\rho (0)}{m_{\chi}} \frac{m}{Am_N} \sqrt{\langle
\upsilon^2\rangle } {\langle \frac{d\Sigma}{du}\rangle }_{dir} 
\label{4.11}  
\eeq
where
\beq
\langle \frac{d\Sigma}{du}\rangle _{dir}=\frac{1}{2 \pi} \int \frac{ 
{\boldmath \upsilon}.\hat{e} H({\boldmath \upsilon}.\hat{e})}
{\sqrt{ \langle \upsilon^2 \rangle}} f({\boldmath \upsilon}, {\boldmath \upsilon}_E)
                       \frac{d\sigma (u,\upsilon )}{du} d^3 {\boldmath \upsilon}
\label{4.12}  
\eeq

\subsection{Directional Differential Event Rate in the Case of Caustic 
Rings.}

The model of Sikivie et al \cite{SIKIVIE}, which is a non isothermal and 
asymmetric one, offers itself as a perfect example for the study of 
directional rates. So we 
are going to begin our discussion with such a case.
 Working as in the previous section we get \cite{JDV99b}
\beq
\langle \frac{d\Sigma}{du} \rangle_{dir}  = \frac{2 \bar{\rho}}
          {\rho(0)}~\frac{a^2}{2 \pi} [\bar {\Sigma} _{S} F_0(u) +
                     \frac{\langle \upsilon ^2 \rangle}{c^2}
                          \bar {\Sigma} _{V} F_1(u) +
                          \bar {\Sigma} _{spin} F_{spin}(u) ]
\label{4.39}  
\eeq
where the $\bar{\Sigma} _{i},i=S,V,spin$ are given by Eqs. (\ref {2.10})-
(\ref {2.10c}). 
The quantities 
$F_0,F_1,F_{spin}$ are obtained from the equations
\beq
F_k(u) = F^2(u)\Psi_k(u)\frac{(1+k)}{2k+1} , k = 0,1
\label{4.40}  
\eeq

\beq
F_{spin}(u) = F_{11}(u) \Psi_0(u) 
\label{4.41}  
\eeq
 In the Sikivie model we find
\barr
\Psi_k(u)& =&\sqrt{\frac{2}{3}}~\sum_{n=1}^N~ \bar{\rho}_n
                   y_n^{2(k-1)}\Theta(\frac{y_n^2}{a^2}-u)
   |(y_{nz}-1-\frac{\delta}{2}~sin\gamma~cos\alpha) {\bf e}_z.{\bf e}  
\nonumber\\
     &+&(y_{ny}+\frac{\delta}{2}~cos\gamma~cos\alpha){\bf e}_y.{\bf e}  
      +(y_{nx}-\frac{\delta}{2}~sin\alpha){\bf e}_x.{\bf e}|  
\label{4.42a}  
\earr
 In the model considered here the z-component of the LSP's velocity
with respect to the galactic center for some rings  is smaller than
sun's velocity, while for some others it
is larger. The components in the y and
the x directions are opposite for the + and - flows. So we will
distinguish the following cases: a) $\hat{e}$ has a
component in the sun's direction of
motion, i.e. $0<\theta < \pi /2$, labeled by u (up). b) Detection
in the direction specified by  $\pi /2 <\theta < \pi $, labeled by 
d (down).  The differential directional rate 
takes a different form depending on which quadrant
the observation is made. Thus :

1. In the first quadrant (azimuthal angle $0 \leq \phi \leq \pi/2)$.
\barr
\langle \frac{dR^i}{du} \rangle & = &\bar{R} \frac{2 \bar{\rho}}{\rho(0)}
    \frac{t}{2 \pi} T(u) [(R^i_z (u) - \cos \alpha~ H^i_1(u))
 |{\bf e}_z.{\bf e}I  
\nonumber \\ 
&+& (R^i_y +cos \alpha H^i_2 (u)+\frac{H^i_c (u)}{2}(|cos\alpha|+cos\alpha))
     |{\bf e}_y.{\bf e} | 
\nonumber \\ 
&+& (R^i_x -sin \alpha H^i_3 (u)+\frac{H^i_s (u)}{2}(|sin\alpha|-sin\alpha))
     |{\bf e}_x.{\bf e} | ]
\label{3.36a}  
\earr
2. In the second quadrant (azimuthal angle $\pi/2 \leq \phi \leq \pi)$
\barr
\langle \frac{dR^i}{du} \rangle & = &\bar{R} \frac{2 \bar{\rho}}{\rho(0)}
    \frac{t}{2 \pi} T(u) [(R^i_z (u) - \cos \alpha~ H^i_1(u)) 
  |{\bf e}_z.{\bf e}| 
\nonumber \\ 
&+& (R^i_y +cos \alpha H^i_2 (u)+\frac{H^i_c (u)}{2}(|cos\alpha|-cos\alpha))
     |{\bf e}_y.{\bf e} | 
\nonumber \\ 
&+& (R^i_x +sin \alpha H^i_3 (u)+\frac{H^i_s (u)}{2}(|sin\alpha|+sin\alpha))
     |{\bf e}_x.{\bf e} | ]
\label{3.37a}  
\earr
3. In the third quadrant (azimuthal angle $\pi \leq \phi \leq 3 \pi/2)$.
\barr
\langle \frac{dR^i}{du} \rangle & = &\bar{R} \frac{2 \bar{\rho}}{\rho(0)}
    \frac{t}{2 \pi} T(u) [(R^i_z (u) - \cos \alpha~ H^i_1(u)) 
  |{\bf e}_z.{\bf e}|  
\nonumber \\ 
&+& (R^i_y -cos \alpha H^i_2 (u)+\frac{H^i_c (u)}{2}(|cos\alpha|-cos\alpha))
     |{\bf e}_y.{\bf e} | 
\nonumber \\ 
&+& (R^i_x +sin \alpha H^i_3 (u)+\frac{H^i_s (u)}{2}(|sin\alpha|+sin\alpha))
     |{\bf e}_x.{\bf e} | ]
\label{3.38a}  
\earr
4. In the fourth quadrant (azimuthal angle $3 \pi/2 \leq \phi \leq 2 \pi)$
\barr
\langle \frac{dR^i}{du} \rangle & = &\bar{R} \frac{2 \bar{\rho}}{\rho(0)}
    \frac{t}{2 \pi} T(u) [(R^i_z (u) - \cos \alpha~ H^i_1(u)) 
       |{\bf e}_z.{\bf e}|  
\nonumber \\ 
&+& (R^i_y -cos \alpha H^i_2 (u)+\frac{H^i_c (u)}{2}(|cos\alpha|-cos\alpha))
     |{\bf e}_y.{\bf e} | 
\nonumber \\ 
&+& (R^i_x -sin \alpha H^i_3 (u)+\frac{H^i_s (u)}{2}(|sin\alpha|-sin\alpha))
     |{\bf e}_x.{\bf e} | ]
\label{3.39a}  
\earr
where $i=u,d$

By the reasoning given above, if one mechanism is dominant,
the parameters 
$R_x,R_y,R_z,H_1,H_2,H_3,H_c,H_s$ for both directions $u$ and $d$
depend only on $\mu_r$ and $a$.  They are all independent
 of the other SUSY parameters.


\subsection{The Directional Differential Event Rate in the Case of 
Velocity Dispersion}
The dependence of the rate depends on the direction of observation
in a rather complicated way. The integrals can only be done 
numerically \cite{Copi99}. To simplify matters we made a power expansion
in $\delta$ and kept terms up to linear in it.
 To make the presentation tractable we will will give expressions
valid only for directions of greatest interest, i.e. close to the 
coordinate axes. In the sun's direction of motion we have a modulated
as well as a non modulated amplitude. In the other two directions we only have
a modulated amplitude. Unlike the case of caustic rings, now the
direction opposite to the sun's direction of motion is favored.
We found it more convenient, however, to present our results as the
absolute value of the difference of the rates in the directions $\hat{e}$
and $-\hat{e}$.

 Working as in the previous subsection we get \cite{JDV99b} 
\beq
\langle \frac{d\Sigma}{du} \rangle_{dir}  = \frac{1}{2 \pi}~a^2
                          [\bar {\Sigma} _{S} F_0(u) +
                     \frac{\langle \upsilon ^2 \rangle}{c^2}
                          \bar {\Sigma} _{V} F_1(u) +
                          \bar {\Sigma} _{spin} F_{spin}(u) ]
\label{4.39b}  
\eeq
The quantities 
$F_0,F_1,F_{spin}$ are obtained from the equations
\beq
F_k(u) = F^2(u)\Psi_k(u)\frac{(1+k)a^2}{2k+1} , k = 0,1
\label{4.40b}  
\eeq
Now
\barr
\Psi_k(u)& = &\frac{1}{2} [(\psi _{(0),k}(a\sqrt{u})+
            0.135 \cos \alpha \psi _{(1),k}(a\sqrt{u}))|{\bf e}_z.{\bf e}|  
\nonumber \\ & - &
            0.117 \cos \alpha \psi _{(2),k}(a\sqrt{u}) |{\bf e}_y.{\bf e}|+  
            0.135 \sin \alpha \psi _{(3),k}(a\sqrt{u}) |{\bf e}_x.{\bf e}|]  
\label{4.42}  
\earr
with
\beq
\psi _{(l),k}(x)= N(y_{esc},\lambda) e^{-\lambda}(e^{-1} \Phi_{(l),k}(x)-
                  exp[-y^2_{esc}] \Phi{'}_{(l),k}(x))  
\label{3.36}  
\eeq
\beq
\Phi_{(l),k}(x) =  \frac{2}{\sqrt{6 \pi}} \int_x^{y_{esc}} dy y^{2k-1}
                     exp {(-(1+\lambda) y^2)})( F_{l}(\lambda,2(\lambda+1)y)+
                   G_{l}(\lambda,y)))
\label{3.37}  
\eeq
\beq
\Phi^{'}_{(l),k}(x) =  \frac{2}{\sqrt{6 \pi}} \int_x^{y_{esc}} dy y^{2k-1}
                     exp {(-\lambda y^2)}) G{'}_{l}(\lambda,y))
\label{3.38}  
\eeq
In the above expressions 
\beq
F_i(\lambda,\chi) =\chi cosh \chi - sinh \chi ~~,~~i=0,2,3 
\label{4.31}  
\eeq
\beq
F_1(\lambda,\chi) \, = \, 2(1-\lambda)\, \Big[ \,(\frac{(\lambda+1)\chi^2}
{4(1-\lambda)} + 1) sinh\, \chi - 
\chi \,cosh \,\chi  \, \Big]
\label{4.32}
\eeq
 The purely asymmetric quantities $G_i$ satisfy
\beq
G_i(0,y)=0,~i=0,4  
\label{4.33}  
\eeq
the qualities $ G^{'}_i(0,y)=0,~i=0,4$  refer to the second term of the 
Velocity distribution and were obtained in an analogous fashion.

 If we consider each mode (scalar, spin vector) separately the directional
rate takes the form
\beq
\langle \frac{dR}{du} \rangle_{dir}  = \bar{R}
 \frac{\rho^{'}(0)}{\rho(0)} \frac{t^0~R^0}{4 \pi} 
                       |(1 + \cos \alpha H_1(u)){\bf e}_z.{\bf e}- 
                       \cos \alpha H_2(u) {\bf e}_y.{\bf e}+  
                           \sin \alpha H_3(u)  {\bf e}_x.{\bf e}|  
\label{3.40}  
\eeq

In other words the directional differential modulated amplitude is described
in terms of the three parameters, $H_l(u)$, l=1,2 and 3. The 
unmodulated amplitude $R^0(u)$ is again normalized to unity. 
The parameter $t^0$ entering Eq. (\ref {3.40}) takes care of whatever 
modifications are needed due to the convolution with the 
LSP velocity  distribution in the presence of the nuclear form factors. 

From Eqs. (\ref {4.39b}) - (\ref {3.40}) we see that if we consider each
mode separately the differential modulation amplitudes $H(l)$ take the form
\beq
H_l(u) =0.135\frac{\psi^{(l)}_{k}(a \sqrt{u})}{\psi ^{(0)}_{k}(a \sqrt{u})}~~,
         ~~l=1,3~~;~~  
H_2(u) =0.117 \frac{\psi^{(2)}_{k}(a \sqrt{u})}{\psi ^{(0)}_{k}(a \sqrt{u})}    
\label{4.45}  
\eeq 
Thus in this case the $H_l$ depend only on $a\sqrt{u}$, which coincides with 
the parameter x of Ref. \cite{Smith}.
This means that $H_l$ essentially depend 
only on the momentum transfer, the reduced mass and the size of the nucleus.
 We note that in
the case $\lambda=0$ we have $H_2=0.117$ and $H_3=0.135$

It is sometimes convenient to use the
quantity $R_l$ rather than $H_l$ defined by
\beq
R_l = R^0 H_l, \,  l=1,2,3.
\label{3.46}  
\eeq
The reason is that $H_l$, being the ratio of two quantities, may appear
superficially large due to the denominator becoming small.

\section{The Total  Non-directional Event Rates}
Integrating Eq. (\ref {3.31a})  we
 obtain for the total non-directional rate in the case of caustic rings
the expression:
\beq
R =  \bar{R}~\frac{2 \bar{\rho}}{\rho(0)}~t~
          [1 - h(a,Q_{min})cos{\alpha})] 
\label{3.55a}  
\eeq
where $Q_{min}$ is the energy transfer cutoff imposed by the detector.
 The modulation is described by the parameter $h$. 
 Similarly integrating Eq. (\ref {3.31b}) we obtain for the total
non-directional rate in our isothermal model as follows:
\beq
R =  \bar{R} \frac{\rho^{'}(0)}{\rho(0)}~t~ 
 [(1 + h(a,Q_{min})cos{\alpha})] 
\label{3.55b}  
\eeq
Note the difference of sign in the definition of the modulation amplitude h 
compared to Eq. (\ref {3.55a}). 
where $Q_{min}$ is the energy transfer cutoff imposed by the detector.
 The modulation can be described in terms of the parameter $h$. 

 The effect of folding
with LSP velocity on the total rate is taken into account via the quantity
$t$. The SUSY parameters have been absorbed in $\bar{R}$. From our 
discussion in the case of differential rate it is clear that strictly
speaking the quantities $t$ and $h$ also depend on the SUSY parameters. They do 
not depend on them, however, if one considers the scalar, spin etc. modes 
separately. 

The meaning of $t$ is clear from the above discussion. We only like to
stress that it is a common practice to extract the LSP nucleon cross 
section from the  the expected experimental event rates in order to 
compare with the SUSY predictions as a function of the LSP mass. In
such analysis the factor $ t$ is omitted. It is clear, however, that,
in going from the data to the cross section, one should
divide by $t$. The results   will be greatly affected for large 
reduced mass.

\section{The Total  Directional Event Rates}

 We will again examine separately the case of caustic rings and the
isothermal models considered above.

\subsection{The Total  Directional Event Rates in the Case of Caustic
Rings}

 Integrating Eqs.
 (\ref {3.36a}) -  (\ref {3.39a}) we obtain:

1. In the first quadrant (azimuthal angle $0 \leq \phi \leq \pi/2)$.
\barr
R^i_{dir} & = &\bar{R} \frac{2 \bar{\rho}}{\rho(0)}
    \frac{t}{2 \pi} [(r^i_z  - \cos \alpha~ h^i_1) |{\bf e}_z.{\bf e}|  
\nonumber \\ 
&+& (r^i_y +cos \alpha h^i_2 +\frac{h^i_c }{2}(|cos\alpha|+cos\alpha))
     |{\bf e}_y.{\bf e} | 
\nonumber \\ 
&+& (r^i_x -sin \alpha h^i_3 +\frac{h^i_s }{2}(|sin\alpha|-sin\alpha))
     |{\bf e}_x.{\bf e} | ]
\label{3.56}  
\earr
2. In the second quadrant (azimuthal angle $\pi/2 \leq \phi \leq \pi)$
\barr
R^i_{dir} & = &\bar{R} \frac{2 \bar{\rho}}{\rho(0)}
    \frac{t}{2 \pi}  [(r^i_z  - \cos \alpha~ h^i_1) |{\bf e}_z.{\bf e}|  
\nonumber \\ 
&+& (r^i_y +cos \alpha h^i_2 (u)+\frac{h^i_c }{2}(|cos\alpha|-cos\alpha))
     |{\bf e}_y.{\bf e} | 
\nonumber \\ 
&+& (r^i_x +sin \alpha h^i_3 +\frac{h^i_s }{2}(|sin\alpha|+sin\alpha))
     |{\bf e}_x.{\bf e} | ]
\label{3.57}  
\earr
3. In the third quadrant (azimuthal angle $\pi \leq \phi \leq 3 \pi/2)$.
\barr
R^i_{dir} & = &\bar{R} \frac{2 \bar{\rho}}{\rho(0)}
    \frac{t}{2 \pi}  [(r^i_z  - \cos \alpha~ h^i_1) |{\bf e}_z.{\bf e}| 
\nonumber \\ 
&+& (r^i_y -cos \alpha h^i_2 (u)+\frac{h^i_c (u)}{2}(|cos\alpha|-cos\alpha))
     |{\bf e}_y.{\bf e} | 
\nonumber \\ 
&+& (r^i_x +sin \alpha H^i_3 +\frac{h^i_s }{2}(|sin\alpha|+sin\alpha))
     |{\bf e}_x.{\bf e} | ]
\label{3.58}  
\earr
4. In the fourth quadrant (azimuthal angle $3 \pi/2 \leq \phi \leq 2 \pi)$
\barr
R^i_{dir} & = &\bar{R} \frac{2 \bar{\rho}}{\rho(0)}
    \frac{t}{2 \pi}  [(r^i_z  - \cos \alpha~ h^i_1) |{\bf e}_z.{\bf e}|  
\nonumber \\ 
&+& (r^i_y -cos \alpha h^i_2 +\frac{h^i_c }{2}(|cos\alpha|-cos\alpha))
     |{\bf e}_y.{\bf e} | 
\nonumber \\ 
&+& (r^i_x -sin \alpha h^i_3 +\frac{h^i_s }{2}(|sin\alpha|-sin\alpha))
     |{\bf e}_x.{\bf e} | ]
\label{3.59 }  
\earr

\subsection{The Total  Directional Event Rates in Isothermal Models}

 We remind the reader that in this case we take the difference of the
rates in two opposite directions.

Integrating Eq. (\ref {4.40}) we obtain
\barr
R_{dir}& = &  \bar{R} [\rho^{'}(0)/(\rho(0)] \, (t^0/4 \pi) \, 
            |(1 + h_1(a,Q_{min})cos{\alpha}) {\bf e}~_z.{\bf e}
\nonumber\\  &-& h_2(a,Q_{min})\, 
cos{\alpha} {\bf e}~_y.{\bf e}
                      + h_3(a,Q_{min})\, 
sin{\alpha} {\bf e}~_x.{\bf e}|
\label{4.55}  
\earr
note that in the above expressions, unlike the case of caustic
rings, the rate is  normalized to $t^0/2$ and not to $t$
 In other words the effect of folding
with LSP velocity on the total rate is taken into account via the quantity
$t^0$. All other SUSY parameters have been absorbed in $\bar{R}$, under the
assumptions discussed above in the case of non-directional rates.

 We see that the modulation of the directional total event rate can
be described in terms of three parameters $h_l$, l=1,2,3. 
 In the special case of $\lambda=0$ we essentially have  one 
parameter, namely $h_1$, since then we have $h_2=0.117$ and $h_3=0.135$.

Given the functions $h_l(a,Q_{min})$ one can plot the the expression in
Eq. (\ref {4.55}) as a function of the phase of the earth $\alpha$. 

\section{Results and Discussion}
\bigskip

The three basic ingredients of our calculation were 
the input SUSY parameters (see sect. 5), a quark model for the nucleon
(see sect. 4) and the structure of the nuclei involved (see sect. 6).
The input SUSY parameters used for the results presented in Tables 1,2 and 3
have been calculated in a phenomenologically allowed 
parameter space (cases \#1, \#2, \#3 of Kane {\it et al}~\cite{ref3} and cases \#4-9
of Castano {\it et al}~\cite{ref3}. Our own SUSY parameters will appear 
elsewhere \cite {Gomez}.

For the coherent part (scalar and vector) we used realistic nuclear
form factors and studied three nuclei, representatives
of the light, medium and heavy nuclear isotopes ($Ca$, $Ge$ and $Pb$).
In Tables 8,9 and 10 we show the results obtained for three different quark
models denoted by A (only quarks u and d) and B, C (heavy quarks
in the nucleon).
\begin{table}[t]  
\caption{  
The quantity $\langle dN/dt\rangle_0=\bar{R}t$ in $y^{-1}Kg^{-1}$
and the modulation  parameter h for the coherent vector and scalar 
contributions in the  cases \#1 - \#3 and for three typical nuclei.
}
\begin{center}
\begin{tabular}{|rl|cc|lrrc|}
\hline
\hline
& & & & & & &   \\
 & & \multicolumn{2}{|c|}{Vector $\,\,\,$ Contribution}  &
     \multicolumn{4}{|c|}{Scalar $\,\,\,$ Contribution}  \\
\hline 
& & & & & & &   \\
& & $\langle dN/dt\rangle_0$ &$h$ &\multicolumn{3}{c}{$\langle 
dN/dt\rangle_0 $}& $h$ \\ 
\hline
& & & & & & &   \\
& \multicolumn{1}{c|}{Case} &$(\times 10^{-3})$ &  &
    \multicolumn{1}{c}{ Model $\,\,$ A }& 
    \multicolumn{1}{c}{ Model $\,\,$ B }& 
    \multicolumn{1}{c}{ Model $\,\,$ C }&  \\ 
\hline
& & & & & & &   \\
  &$\#1$& 0.264 &0.029 &$0.151\times 10^{-3}$ &  0.220 &  0.450 &-0.002 \\
Pb&$\#2$& 0.162 &0.039 &$0.410\times 10^{-1}$ &142.860 &128.660 & 0.026 \\
  &$\#3$& 0.895 &0.038 &$0.200\times 10^{-3}$ &  0.377 &  0.602 &-0.001 \\
\hline
& & & & & & &   \\
  &$\#1$& 0.151 &0.043 &$0.779\times 10^{-4}$ &  0.120 &  0.245 & 0.017 \\
Ge&$\#2$& 0.053 &0.057 &$0.146\times 10^{-1}$ & 51.724 & 46.580 & 0.041 \\
  &$\#3$& 0.481 &0.045 &$0.101\times 10^{-3}$ &  0.198 &  0.316 & 0.020 \\
\hline
& & & & & & &   \\
  &$\#1$& 0.079 &0.053 &$0.340\times 10^{-4}$ &  0.055 &  0.114 & 0.037 \\
Ca&$\#2$& 0.264 &0.060 &$0.612\times 10^{-2}$ & 22.271 & 20.056 & 0.048 \\
  &$\#3$& 0.241 &0.053 &$0.435\times 10^{-4}$ &  0.090 &  0.144 & 0.038 \\
\hline
\hline
\end{tabular}
\end{center}
\end{table}
\begin{table}[t]  
\caption{  
The spin contribution in the $LSP-^{207}Pb$ scattering 
for two cases: EMC data and NQM Model for solutions $\#1, \#2, \#3$. 
}
\begin{center}
\begin{tabular}{|l|ll|lc|}
\hline
\hline
& & & &   \\
& \multicolumn{2}{|c}{\hspace{1.2cm}EMC \hspace{.2cm} DATA} \hspace{.8cm} &
 \multicolumn{2}{|c|}{\hspace{1.2cm}NQM \hspace{.2cm} MODEL} \\ 
\hline
& & & &   \\
Solution & \hspace{.2cm}$\langle dN/dt \rangle _0$ $(y^{-1} Kg^{-1})$
  \hspace{.2cm} & $ h $ & 
$\hspace{.2cm} \langle dN/dt \rangle_0 $ $ (y^{-1} Kg^{-1})$ & $ h $ \\ 
\hline
& & & &   \\
$\#1  $  &$0.285\times 10^{-2}$& 0.014 &$0.137\times 10^{-2}$& 0.015  \\
$\#2  $  & 0.041               & 0.046 &$0.384\times 10^{-2}$& 0.056  \\
$\#3  $  & 0.012               & 0.016 &$0.764\times 10^{-2}$& 0.017  \\
\hline
\hline
\end{tabular}
\end{center}
\end{table}
\begin{table}[t]  
\caption{  
The same quantities as in Table 8 in the case of $Pb$ for the solutions 
 $\#4-\#9$. $\#8$ snd $\#9$ are no-scale models. The values of 
$\langle dN/dt \rangle_0=\bar{R}t$ for Model A and the Vector part 
must be multiplied by $\times 10^{-2}$.
}
\begin{center}
\begin{tabular}{|l|crrr|lr|lll|}
\hline
\hline
 & & & & & & & & &   \\
  & \multicolumn{4}{|c|}{Scalar$\,$ Part} & 
    \multicolumn{2}{c|}{Vector $\,$ Part} &
    \multicolumn{3}{c|}{Spin   $\,$ Part} \\
\hline 
 & & & & & & & & &   \\
& \multicolumn{3}{c}{$\big<\frac{dN}{dt}\big>_0$}& $h$ & 
$\big<\frac{dN}{dt}\big>_0$ & $h$ & \multicolumn{2}{c}
{$\big<\frac{dN}{dt}\big>_0$} & $h$ \\ 
\hline
 & & & & & & & & &   \\
Case & A & B & C & & & & EMC & NQM & \\ 
\hline
 & & & & & & & & &   \\
$\#4 $& 0.03 & 22.9& 8.5 & 0.003& 0.04 & 0.054
 & $0.80\, 10^{-3}$& $0.16\, 10^{-2}$& 0.015 \\
$\#5 $& 0.46 & 1.8& 1.4 & -0.003& 0.03 & 0.053
 & $0.37\, 10^{-3}$& $0.91\, 10^{-3}$& 0.014 \\
$\#6 $& 0.16 & 5.7& 4.8 & 0.007& 0.11 & 0.057
 & $0.44\, 10^{-3}$ & $0.11\, 10^{-2}$& 0.033 \\
$\#7 $&  4.30 & 110.0& 135.0 & 0.020& 0.94 & 0.065
 & 0.67 & 0.87 & 0.055 \\
$\#8 $& 2.90 & 73.1& 79.8 & 0.020& 0.40 & 0.065
 & 0.22 & 0.35 & 0.055 \\
$\#9 $&  2.90 & 1.6& 1.7 & 0.009& 0.95 & 0.059
 & 0.29 & 0.37 & 0.035 \\
\hline
\hline
\end{tabular}
\end{center}
\end{table}
We see that the results vary substantially and are very sensitive to 
the presence of quarks other than u and d into the nucleon.
The spin contribution, arising from the axial current,
was computed in the case of  a number of both light and heavy
nuclei, including the $^{207}Pb$ system. 
For the isovector axial coupling the transition from the quark to
the nucleon level is trivial (a factor of $g_A=1.25)$. For the
isoscalar axial current we considered two possibilities 
depending on the portion of the nucleon spin which is attributed to the
quarks, indicated by EMC and NQM.~\cite{KVprd}
The ground state wave function of $^{208}Pb$ was obtained
by diagonalizing the nuclear Hamiltonian~\cite{KV90}-\cite{Kuo} 
in a 2h-1p space which is standard for this doubly magic nucleus.   
The momentum dependence of the matrix elements was taken into account and all 
relevant multipoles were retained (here only monopole and quadrupole).

In  Table 5, we compare the spin matrix elements at $q=0$ for the 
most popular targets considered for LSP detection $^{207}Pb$, 
$^{73}Ge$, $^{19}F$,$^{23}Na$ and $^{29}Si$.
 We see that, even though the spin matrix elements $\Omega^2$ in the
case of $^{207}Pb$ are about factor of three smaller than those for 
$^{73}Ge$ obtained 
in Ref.~\cite{Ress} (see Table 5), their contribution to the total
cross section is almost the same (see Table 6) for LSP masses around 
$100 \, GeV$. Our final results for the quark models (A, B, C, NQM, EMC)
are presented in Tables 8, 9 for SUSY models \#1-\#3 of Kane {\it et al}
~\cite{ref3}
and Table 10 for SUSY models \#4-\#9 of Castano {\it et al}~\cite{ref3}

 In discussing the effects of folding with the LSP velocity combined with the
nuclear form factor and specialized our results for the target $^{127}I$.  To
this end we considered only the scalar interaction and studied 
the effects of the detector energy cutoffs, by considering two  typical
cases $Q_{min}=10,~20$ KeV.  

Special attention was paid to the the directional rates and the  
modulation effect due to the annual motion of the earth.

We will start our discussion with the non isothermal velocity spectrum
due to  caustic rings resulting from the self-
similar model of Sikivie et al \cite{SIKIVIE}. 
 
 The total rates are 
described in terms of the quantities $t,r^i_x,r^i_y,r^i_z$ for the
unmodulated amplitude  and $h,h^i_1,h^i_2,h^i_3,h^i_c,h^i_s$ $i=u,d$
for the modulated one. 
In Table 11 I we show how these quantities vary with the detector energy cutoff
and the LSP mass. Of the above list only the quantities $t$ and $h$
enter the non-directional rate. We notice that the usual
modulation amplitude $h$ is smaller than the one arising in isothermal
models \cite {JDV99,JDV99b}. The reason is that there are cancelations among
the various rings, since some rings are characterized by
$y_{nz}>1$, while for some others $y_{nz}<1$ (see Table 7). Such
cancelations are less pronounced in the isothermal models.
 As expected, the parameter t, which contains the 
effect of the nuclear form factor and the LSP velocity dependence,
decreases as the reduced mass increases.
 
  In the case of isothermal models we will limit ourselves to the discussion
of the directional rates. In the special case of the direction of observation 
being close to the coordinate axes the rate is described in terms of the three
quantities $t_0$ and $h_i,~i=1,2,3$ (see Eq.  (\ref{4.55})). 
These are shown in tables 12-14 for various values of $Q_{min}$ and $\lambda$. For the differential rate the reader is 
referred to our previous work \cite {JDV99,JDV99b}.
 We mention again that $h_2$ and $h_3$ 
are constant, 0.117 and 0.135 respectively, in the symmetric case. On the other
hand $h_1,h_2$ and $h_3$ substantially increase in the presence of asymmetry. 

\begin{table}[t]  
\caption{The quantities $t$ and $h$ entering the total non-directional
rate in the case of the
target $_{53}I^{127}$ for various LSP masses and $Q_{min}$ in KeV. 
Also shown are the quantities $r^i_j,h^i_j$
 $i=u,d$ and $j=x,y,z,c,s$, entering the directional rate for no energy
cutoff. For definitions see text. 
}
\begin{center}
\footnotesize
\begin{tabular}{|c|c|rrrrrrr|}
\hline
\hline
& & & & & & & &     \\
&  & \multicolumn{7}{|c|}{LSP \hspace {.2cm} mass \hspace {.2cm} in GeV}  \\ 
\hline 
& & & & & & & &     \\
Quantity &  $Q_{min}$  & 10  & 30  & 50  & 80 & 100 & 125 & 250   \\
\hline 
& & & & & & & &     \\
t      &0.0&1.451& 1.072& 0.751& 0.477& 0.379& 0.303& 0.173\\
h      &0.0&0.022& 0.023& 0.024& 0.025& 0.026& 0.026& 0.026\\
$r^u_z$&0.0&0.726& 0.737& 0.747& 0.757& 0.760& 0.761& 0.761\\
$r^u_y$&0.0&0.246& 0.231& 0.219& 0.211& 0.209& 0.208& 0.208\\
$r^u_x$&0.0&0.335& 0.351& 0.366& 0.377& 0.380& 0.381& 0.381\\
$h^u_z$&0.0&0.026& 0.027& 0.028& 0.029& 0.029& 0.030& 0.030\\
$h^u_y$&0.0&0.021& 0.021& 0.020& 0.020& 0.019& 0.019& 0.019\\
$h^u_x$&0.0&0.041& 0.044& 0.046& 0.048& 0.048& 0.049& 0.049\\
$h^u_c$&0.0&0.036& 0.038& 0.040& 0.041& 0.042& 0.042& 0.042\\
$h^u_s$&0.0&0.036& 0.024& 0.024& 0.023& 0.023& 0.022& 0.022\\
$r^d_z$&0.0&0.274& 0.263& 0.253& 0.243& 0.240& 0.239& 0.239\\
$r^d_y$&0.0&0.019& 0.011& 0.008& 0.007& 0.007& 0.007& 0.007\\
$r^d_x$&0.0&0.245& 0.243& 0.236& 0.227& 0.225& 0.223& 0.223\\
$h^d_z$&0.0&0.004& 0.004& 0.004& 0.004& 0.004& 0.004& 0.004\\
$h^d_y$&0.0&0.001& 0.000& 0.000& 0.000& 0.000& 0.000& 0.000\\
$h^d_x$&0.0&0.022& 0.021& 0.021& 0.020& 0.020& 0.020& 0.020\\
$h^d_c$&0.0&0.019& 0.018& 0.018& 0.017& 0.017& 0.017& 0.017\\
$h^d_s$&0.0&0.001& 0.001& 0.000& 0.000& 0.000& 0.000& 0.000\\
\hline 
& & & & & & & &     \\
t    &10.0&0.000& 0.226& 0.356& 0.265& 0.224& 0.172& 0.098\\
h    &10.0&0.000& 0.013& 0.023& 0.025& 0.025& 0.026& 0.026\\
\hline 
& & & & & & & &     \\
t    &20.0&0.000& 0.013& 0.126& 0.139& 0.116& 0.095& 0.054\\
h    &20.0&0.000& 0.005& 0.017& 0.024& 0.025& 0.026& 0.026\\
\hline
\hline
\end{tabular}
\end{center}
\end{table}
\begin{table}[t]  
\caption{The quantities $t^{0},h_1$ and $h_m$ for $\lambda=0$ in the case of the
target $_{53}I^{127}$ for various LSP masses and $Q_{min}$ in KeV (for
definitions see text). Only the scalar contribution is considered. Note that in
this case $h_2$ and $h_3$ are constants equal to 0.117 and 0.135 respectively.}
\begin{center}
\footnotesize
\begin{tabular}{|l|c|rrrrrrr|}
\hline
\hline
& & & & & & & &     \\
&  & \multicolumn{7}{|c|}{LSP \hspace {.2cm} mass \hspace {.2cm} in GeV}  \\ 
\hline 
& & & & & & & &     \\
Quantity &  $Q_{min}$  & 10  & 30  & 50  & 80 & 100 & 125 & 250   \\
\hline 
& & & & & & & &     \\
$t^0$ &0.0&1.960&1.355&0.886&0.552&0.442&0.360&0.212 \\
$h_1$ &0.0&0.059&0.048&0.037&0.029&0.027&0.025&0.023 \\
\hline 
& & & & & & & &     \\
$ t^0$ &10.&0.000&0.365&0.383&0.280&0.233&0.194&0.119 \\
$h_1$ &10.&0.000&0.086&0.054&0.038&0.033&0.030&0.025 \\
\hline 
& & & & & & & &     \\
$t^0$ &20.&0.000&0.080&0.153&0.136&0.11&0.102&0.065 \\
$h_1$ &20.&0.000&0.123&0.073&0.048&0.041&0.036&0.028 \\
\hline
\hline
\end{tabular}
\end{center}
\end{table}
\begin{table}[t]  
\caption{The same as in the previous table, but for the value of the asymmetry 
parameter $\lambda=0.5$.}
\begin{center}
\footnotesize
\begin{tabular}{|l|c|rrrrrrr|}
\hline
\hline
& & & & & & & &     \\
&  & \multicolumn{7}{|c|}{LSP \hspace {.2cm} mass \hspace {.2cm} in GeV}  \\ 
\hline 
& & & & & & & &     \\
Quantity &  $Q_{min}$  & 10  & 30  & 50  & 80 & 100 & 125 & 250   \\
\hline 
& & & & & & & &     \\
$t^0$ & 0.0 &2.309&1.682&1.153&0.737&0.595&0.485&0.288 \\
$h_1$ & 0.0 &0.138&0.128&0.117&0.108&0.105&0.103&0.100\\
$h_2$ & 0.0 &0.139&0.137&0.135&0.133&0.133&0.133&0.132\\
$h_3$ & 0.0 &0.175&0.171&0.167&0.165&0.163&0.162&0.162\\
\hline 
& & & & & & & &     \\
$t^0$ & 10. &0.000&0.376&0.468&0.365&0.308&0.259&0.160\\
$h_1$ & 10. &0.000&0.174&0.139&0.120&0.114&0.110&0.103\\
$h_2$ & 10. &0.000&0.145&0.138&0.135&0.134&0.134&0.133\\
$h_3$ & 10. &0.000&0.188&0.174&0.167&0.165&0.164&0.162\\
\hline 
& & & & & & & &     \\
$t^0$ & 20. &0.000&0.063&0.170&0.171&0.153&0.134&0.087\\
$h_1$ & 20. &0.000&0.216&0.162&0.133&0.124&0.118&0.107\\
$h_2$ & 20. &0.000&0.155&0.143&0.137&0.136&0.135&0.133\\
$h_3$ & 20. &0.000&0.209&0.182&0.171&0.168&0.166&0.164\\
\hline
\hline
\end{tabular}
\end{center}
\end{table}
\begin{table}[t]  
\caption{The same as in the previous, but for the value of the asymmetry parameter
 $\lambda=1.0$.}
\begin{center}
\footnotesize
\begin{tabular}{|l|c|rrrrrrr|}
\hline
\hline
& & & & & & & &     \\
&  & \multicolumn{7}{|c|}{LSP \hspace {.2cm} mass \hspace {.2cm} in GeV}  \\ 
\hline 
& & & & & & & &     \\
Quantity &  $Q_{min}$  & 10  & 30  & 50  & 80 & 100 & 125 & 250   \\
\hline 
& & & & & & & &     \\
$t^0$ & 0.0 &2.429&1.825&1.290&0.837&0.678&0.554&0.330\\
$h_1$ & 0.0 &0.192&0.182&0.170&0.159&0.156&0.154&0.150 \\
$h_2$ & 0.0 &0.146&0.144&0.141&0.139&0.139&0.138&0.138\\
$h_3$ & 0.0 &0.232&0.222&0.211&0.204&0.202&0.200&0.198\\
\hline 
& & & & & & & &     \\
$t^0$ & 10. &0.000&0.354&0.502&0.410&0.349&0.295&0.184\\
$h_1$ & 10. &0.000&0.241&0.197&0.174&0.167&0.162&0.154\\
$h_2$ & 10. &0.000&0.157&0.146&0.142&0.140&0.140&0.138\\
$h_3$ & 10. &0.000&0.273&0.231&0.213&0.208&0.205&0.200\\
\hline 
& & & & & & & &     \\
$t^0$ & 20. &0.000&0.047&0.169&0.186&0.170&0.150&0.100\\
$h_1$ & 20. &0.000&0.297&0.226&0.190&0.179&0.172&0.159\\
$h_2$ & 20. &0.000&0.177&0.153&0.144&0.142&0.141&0.139\\
$h_3$ & 20. &0.000&0.349&0.256&0.224&0.216&0.211&0.203\\
\hline
\hline
\end{tabular}
\end{center}
\end{table}
\section{Conclusions}
\bigskip
In the present paper we have calculated the parameters, which described
the event rates for direct detection of SUSY dark matter. We found that
the event rates are quite small and only in a small segment of the 
allowed parameter space they are above the present experimental goals.
 One, therefore, is looking for
characteristic signatures, which will aid the experimentalists in reducing
background. These are two: a) The non directional event rates, which are
correlated with the motion of the Earth (modulation effect) and b)
 The directional event rates,
which are correlated with both the velocity of the sun, around the center of
 the Galaxy, and the velocity of the Earth.
 We separated our discussion into
two parts. The first deals with the elementary aspects, SUSY parameters and
nucleon structure, and is given in terms of the nucleon cross-section. The
second deals with the transition from the nucleon to the nuclear level. 
In the second step we also studied the dependence of the rates on the
energy cut off imposed by the detector. We presented our results in a 
fashion understandable by the
 experimentalists. We specialized our results in the case of
the coherent process in the case of $^{127}I$, but
we expect the conclusions to be quite general.

The needed local density and velocity spectrum of the LSP were 
obtained in two special classes: 1)  Non isothermal models and 
2) Isothermal models.
 As we have already mentioned the actual situation may be a combination of an
isothermal contribution and late infall of dark matter.
In the present treatment we consider each of these distributions
separately.

In the first case case we assumed  a late in-fall of dark matter into 
our 
galaxy.  the needed parameters were taken from the work of Sikivie
 et al \cite{SIKIVIE} in the context of a self-similar
model, which yields 40 caustic rings.
 Our results, in particular the parameters $t$, see Table 11, 
indicate that for large reduced mass, the kinematical advantage of
$\mu _r$ (see Eqs.
(\ref{2.10})- (\ref{2.10c}) is partly lost when the nuclear form factor and the 
convolution with the velocity distribution are taken into account. 
Also, if one  attempts to extract the LSP-nucleon cross section from
the data, in order to compare with the predictions of SUSY models,
one must take $t$ into account, since for large reduced mass $t$ is
different from unity.

In the case of the non-directional total event rates we find that 
the maximum no longer
occurs around June 2nd, but about six months later. The difference
between the maximum and the minimum is about $4\%$, smaller
than that predicted by the symmetric isothermal models \cite{JDV99,JDV99b}. 
In the case of the directional rate we found that the rates depend on the 
direction of observation. The biggest rates are obtained, If the
observation is made close to the direction of the sun's motion.
The directional rates are suppressed compared to the usual 
non-directional rates by the factor 
$f_{red}=\kappa/(2 \pi)$. We find that $\kappa=r^u_z \simeq 0.7$, 
if the observation is made in the sun's direction of motion, while
$\kappa\simeq 0.3$ in the opposite direction.
The modulation is a bit larger than in the non-directional case, but the
largest value, 
$8\%$, is not obtained along the sun's direction of motion, but 
in the x-direction (galactocentric direction).

 In the case of the isothermal models we restricted our discussion to the 
directional event rates. The reduction factor of the total directional
rate, along the sun's direction of motion, compared to the total non
directional rate depends, of course, on the
nuclear parameters, the reduced mass and the asymmetry parameter $\lambda$
\cite{JDV99b}.
It is given by the parameter $f_{red}=t_0/(4 \pi~t)=\kappa/(2 \pi)$. We find
that $\kappa$ is around 0.6 for no asymmetry and around 0.7 for maximum
asymmetry ($\lambda=1.0$). In other words it is not very different from the 
naively expected $f_{red}=1/( 2 \pi)$, i.e.  $\kappa=1$. 
The modulation of the directional rate 
increases with the asymmetry parameter $\lambda$ and it depends, of
course, on
 the direction of observation. For $Q_{min}=0$ it can reach values up 
to $23\%$. Values up to $35\%$ are possible for large values of
 $Q_{mon}$, but they occur at the expense of the total
number of counts.
 In all cases our results, in particular the parameters $t$, see Table 11, 
 and $t_0$, see Tables 12-14, 
indicate that for large reduced mass, the kinematical advantage of
$\mu _r$ (see Eqs.
(\ref{2.10})- (\ref{2.10c}) is partly lost when the nuclear form factor and the 
convolution with the velocity distribution are taken into account. 
To be more precise, if one  attempts to extract the LSP-nucleon cross section 
from the data, in order to compare with the predictions of SUSY models,
one must take $t$ into account, since for large reduced mass $t$ is
different from unity.
\bigskip
{\it Acknowledgments:} Partial support by  TMR No ERB FMAX-CT96-0090
of the European Union is happily acknowledged.

\end{document}